\documentclass{ws-ijmpd}
\usepackage[super,compress]{cite}

\usepackage[usenames,dvipsnames]{color}

\newcommand{\map}[1]{\ensuremath{\mathcal{M}_{\rm #1}}}

\let\trimmarks\relax

\begin{document}

\markboth{B\' ela Szil\' agyi}
{Key Elements of Robustness in Binary Black Hole Evolutions using Spectral Methods}

%
\catchline{}{}{}{}{}
%

\title
{Key Elements of Robustness in Binary Black Hole Evolutions using Spectral Methods}

\author{B\' ela Szil\' agyi}

\address{Theoretical Astrophysics 350-17, \\
    California Institute of Technology, \\ Pasadena, CA 91125, USA\\
bela@caltech.edu}

\maketitle

\begin{history}
\received{24 April 2014}
\accepted{26 April 2014}
\end{history}

\begin{abstract}
As a network of advanced-era gravitational wave detectors is nearing its design
sensitivity, efficient and accurate waveform modeling becomes more and more relevant.
Understanding of the nature of the signal being sought can have an order unity
effect on the event rates seen in these instruments.   The paper provides a description
of key elements of the Spectral Einstein Code ({\tt SpEC}), 
with details of our spectral adaptive mesh
refinement (AMR) algorithm that has been optimized for binary black hole (BBH) evolutions.  
We expect that the gravitational
 waveform catalog produced by our code will have a central importance in both the detection
and parameter estimation of gravitational waves in these instruments.

\end{abstract}


\ccode{PACS numbers: 04.25.dg,04.30.-w}


\section{Introduction}

A fundamental consequence of the field equations written down by Einstein, in his Theory of General Relativity,
is the existence of singular solutions that are causally disconnected from  remote observers.  That is, surrounding
the singularity will be an event horizon defined as the boundary of the region from within which a physical observer
cannot escape.   Astrophysical observations have lead to the realization that these objects are not only of theoretical
importance. They are more than just the ``point charge solution'' of Einstein's field equations. In fact, they are key players
in how our universe functions.   Given the large amount of observational evidence for the existence of black holes,
it is a very natural question  to ask -- will these objects ever collide?  If yes, what measurable quantities can be used
to identify such an event.  We know that galaxies collide (ours being one such ``collided'' galaxy).  We know that galaxies
host super massive black holes.  Therefore black holes must collide as well.

The intent of this paper is to highlight some of the key elements of a particular
numerical relativity code, the { Sp}ectral { E}instein { C}ode ({\tt SpEC}),
that allowed it to reach ``production-level'' in simulating binary black hole (BBH) mergers.

It is important to emphasize at a very early point in the paper that, 
though the current manuscript has a single author, the {\tt SpEC} code is the result of years of work
of a large number of people within what is known as the ``SXS collaboration''.
While a fair portion of the paper will be describing code contributions of the author,  
we will also make an effort to give due credit to all  others involved. 

Throughout the paper we will use Latin indexes  $a,b,...$ from the beginning of the alphabet for space-time quantities, 
while $i,j,k,...$ will stand for spatial indexes.     Partial derivatives will be denoted by $\partial_a f$ 
while covariant derivatives will be written as $\nabla _a f$.

\section{Evolution System}

\subsection{The Generalized Harmonic System}

Consider a spacetime metric tensor $\psi_{ab}$,
\begin{equation}
ds^2 = \psi_{ab} dx^a  dx^b .
\end{equation}
The associated Christoffel symbol $\Gamma_{abc}$ is defined as
\begin{equation}
\Gamma_{abc} = \frac{1}{2}\left( \partial_b \psi_{ac} + \partial_c \psi_{ab} - \partial_a \psi_{bc} \right) .
\label{eq:GammaDef}
\end{equation}
We will refer to the trace of the Christoffel symbol as 
\begin{equation}
\Gamma_a \equiv \psi^{bc} \Gamma_{abc}  .
\label{eq:TrGammaDef}
\end{equation}

The Einstein Equations, in a concise form, can be written as
\begin{equation}
G_{ab} = R_{ab} - \frac{1}{2} \psi_{ab} R = 0
\end{equation}
where $R=\psi^{ab} R_{ab}$ is the trace of the Ricci tensor $R_{ab}$ which, in turn, is given by
\begin{equation}
R_{ab} = - \frac{1}{2} \psi^{cd} \partial_c \partial_d \psi_{ab} 
+ \nabla_{(a} \Gamma_{b)} 
+ \psi^{cd} \psi^{ef} \left(\partial_e \psi_{ca} \partial_f \psi_{db} - \Gamma_{ace} \Gamma_{bdf}\right) .
\label{eq:Ricci}
\end{equation}

An essential part of being able to numerically evolve a space time is to establish a well-posed initial boundary value problem.  
``Well-posed'' here translates into saying that given some plausible initial
and boundary data a unique solution will exist and that small perturbations in the freely specifiable data
(on the initial slice or on the
boundary) will result in small changes of the solution to the system under consideration.  One approach to establish well-posedness
of a system of partial differential equations is to show that the principal part (the terms containing the highest order derivatives)
can be written as a first order symmetric hyperbolic system.

A careful look at the right-hand side (RHS) of Eq.~(\ref{eq:Ricci}) shows that the principal part consists of two
terms: the wave operator acting on the metric and derivatives of $\Gamma_b$ (which itself consists of first derivatives
of the metric).   This, together with the identity
\begin{equation}
\psi_{ab}\Delta^c \Delta_c x^a = - \Gamma_b
\end{equation}
leads to the idea that one can think of $\Gamma_b$ as freely specifiable gauge freedom\cite{Friedrich1985}. That is, rather than assigning
values $x^a$ to each point of the manifold in some arbitrary (but smooth) way, one can think of the coordinates being determined
indirectly through a wave equation
\begin{equation}
 \psi_{ab}\Delta^c \Delta_c x^b = H_a
\end{equation}
and regarding $H_a$ as freely specifiable.
Specifying $H_a$, with suitable initial and boundary conditions, is equivalent to specifying $x^a$.  
Substituting $H_a = -\Gamma_a$ into Eq.~(\ref{eq:RicciH}) gives
\begin{equation}
R^{H}_{ab} = - \frac{1}{2} \psi^{cd} \partial_c \partial_d \psi_{ab} 
+ \nabla_{(a} H_{b)} 
+ \psi^{cd} \psi^{ef} \left(\partial_e \psi_{ca} \partial_f \psi_{db} - \Gamma_{ace} \Gamma_{bdf}\right) .
\label{eq:RicciH}
\end{equation}
If the gauge source function $H_a$ is prescribed as a function of the coordinates $(x^a)$ and the metric $\psi_{ab}$ but does not 
depend on derivatives of the metric, 
\begin{equation}
H_a = F_a\left(x^c,\psi_{cd}\right) ,
\end{equation}
then the sole term of the principal part of $R^{H}_{ab}$ consists of the wave operator acting
on the metric.   This leads to a well-posed system, known as the generalized harmonic formulation of  Einstein's equations.

The identification  $H_a = -\Gamma_a$ induces a constraint in the generalized harmonic system, as $H_a$ is now a free function,
while the trace of the Christoffel 
symbol is determined by derivatives of the metric resulting form the evolution equations.  These equations are equivalent
to the Einstein system only in the limit in which the constraint ${\cal C}_a = H_a + \Gamma_a$ vanishes.   It has been shown\cite{Friedrich2005} that
as a consequence of the Bianchi identities, the constraints propagate according to their own set of governing equations,
\begin{equation}
0 = \Delta^b \Delta_b {\cal C}_a + {\cal C}^b \Delta_{(a} {\cal C}_{b)} .
\label{eq:ConstraintProp}
\end{equation}
Thus the constraints themselves propagate as a set of coupled scalar waves implying that
 their evolution system is well-posed as well.  This in turn means that small perturbations of the constraints in either the initial
data or boundary data will not result in uncontrollable runaway solutions of the constraint system.  In addition, the form of the source term of Eq.~(\ref{eq:ConstraintProp}) tells us that for vanishing initial and boundary data on the constraints, they will
remain zero in the entire evolved spacetime region.

In the actual numerical simulations constraint violating modes are generated at each time-step and on a variety of
length-scales.  Well-posedness is essential but not sufficient, as it does not exclude exponentially growing modes.
Constraint propagation can further be improved by adding what is known as constraint damping terms to the evolution system.
Given that the constraints are formed of first derivatives of the metric, one can add arbitrary combination of these
 (but not their derivatives) to the evolution system without modifying its principal part.   
A number of such terms have been proposed in the 
literature.\cite{ Brodbeck1999 ,Gundlach2005,Pretorius2005a,Babiuc2006}
The one employed by  the {\tt SpEC} code can be written as

\begin{equation}
0 = R_ab - \nabla_{(a} {\cal C}_{b)} + \gamma_0 \left[ t_{(a} {\cal C}_{b)} - \frac{1}{2} \psi_{ab} t^c {\cal C}_c \right] ,
\end{equation}
where $t_a$ is the future directed unit timelike normal to the $t=$constant surfaces of the spacetime manifold, while
$\gamma_0$ is a free parameter.\footnote{In a typical BBH simulation we set $\gamma_0$ to be a smooth function
of coordinates, with $\gamma_0=O(10)$ in the inner region of the simulation and $\gamma_0 = 10^{-3}$ near the outer
boundary of the computational domain.}

\subsection{Damped Harmonic Gauge}
\label{sec:DampedHarmonicGauge}

Next we turn our attention to various choices of the gauge source function $H_a$.   One immediate choice is to set the components
of $H_a$  to
zero, resulting in what is known as the harmonic gauge.  Substituting this into
Eq.~(\ref{eq:RicciH}) leads to the harmonic formulation of Einstein's equations.  Harmonic coordinates determine
the fields $(t,x,y,z)$ by a wave equation and suitable initial and boundary values. 
In flat space this is feasible, as the `natural' coordinates $(t,x,y,z)$ are trivial solutions of the wave operator,
\begin{equation}
\left( -\partial_t^2 + \partial_x^2 + \partial_y^2 + \partial_z^2 \right) \left( x^a  \right) = 0
\end{equation}
In a non-trivial space time, however, in the presence of an actual gravitational potential, the scalar wave operator may
amplify fields propagating under its action and form singularities.  The net effect would be a singular coordinate system
(with metric coefficients that become very large).  Better choices are needed.

In his ground-breaking work, Pretorius \cite{Pretorius2005a}  was able to use harmonic spatial coordinates but had to
prescribe the time-component $H_t$ of the gauge source function by a damped wave equation that
prevented the lapse function from collapsing to zero (which would be a signature of a singular time-coordinate).  For generic binary black hole mergers, we found that better gauge choices are needed.

The current favorite gauge condition for binary black hole systems evolved with {\tt SpEC} is the damped harmonic 
gauge\cite{Lindblom2009c,Szilagyi:2009qz}, described by
\begin{equation}
H_a =  \mu_L \log\left(\frac{\sqrt{g}}{N} \right) t_a - \mu_S N^{-1} g_{ai} N^i
\label{eq:InertialDampedHarmonicGauge}
\end{equation}
\begin{equation}
g_{ab} = \psi_{ab} + t_a t_b = \psi_{ab} + N^2 \delta_a^t \delta_b^t ,
\end{equation}
where $g$ is the determinant of the spatial 3-metric $g_{ij}$,
 $N$ is the lapse function and $N^i$ is the shift, defined as
\begin{equation}
N = (- \psi^{tt})^{-1/2}, \quad N^i = - \psi^{it} / \psi^{tt} .
\end{equation}
The coefficients $\mu_L, \mu_S$ are used to control the amount of damping.   We find that the choice
\begin{equation}
\mu_S = \mu_L = \mu_0 \left[ \log \left( \frac{\sqrt{g}}{N} \right) \right]^2
\end{equation}
works well, where $\mu_0$ is a time-dependent coefficient that is rolled from zero to one in  a smooth
time-dependent way at the start of the simulation in order to reduce numerical error induced
by  initial gauge-dynamics of the  BBH system.

Gauge conditions are intended to impose  a condition on coordinates and, as such, they cannot be coordinate-independent
(or covariant). This holds for our system as well.
The form given in Eq.~(\ref{eq:InertialDampedHarmonicGauge}) acts a damping condition on the 
inertial frame\footnote{We define
{\em inertial} frame the coordinate system in which the evolution equations are evolved. In these
coordinates the far-field metric is a perturbation around Minkowski, while the coordinate position
of the individual objects is time-dependent, as the black holes orbit each other. In a more precise sense, the 
inertial coordinates $(t,x^i)$ are 
determined by the initial and boundary data of the evolved metric quantities, 
along with our choice of harmonic gauge source functions $H_a$
defines the coordinates. }
damps the value of $\log\left(\frac{\sqrt{g}}{N} \right)$ towards small values, effectively preventing the lapse from collapsing and/or the 
metric volume density $\sqrt{g}$ from becoming very large.

The drawback of this gauge choice is that near merger, when objects tend to have larger coordinate velocities, it results in
distortion.   An alternative would be to impose the same condition in a comoving frame\footnote{The {\em comoving}
frame, defined in Eq.~(\ref{eq:ComovingFrame}), is
the frame connected to the inertial frame by  translation, rotation and scaling.  In this frame the
coordinate centers of the
individual black holes are time-independent throughout the inspiral and plunge of the binary.  
Definition of a such frame
is essential in our code as the underlying numerical grid is rigid, with fixed excision boundaries.  These excision
surfaces are kept just inside the dynamically evolved apparent horizons by the sequence of maps
given in Eq.~(\ref{eq:MapSequence}).} (with coordinates
$\tilde x^{\tilde k}$) of the binary.

When doing
so, one has to be mindful of the fact that $H_a$ in general has transformation properties that depend on the particular choice of
gauge condition (in our case Eq.~(\ref{eq:InertialDampedHarmonicGauge})), 
while $\Gamma_a$ transforms as prescribed by its definition (see Eqs.~(\ref{eq:GammaDef})-(\ref{eq:TrGammaDef})). 
 As a consequence, naively substituting the co-moving
frame metric quantities $\tilde g, \tilde g_{\tilde a \tilde i}, \tilde N^{\tilde i}, \tilde t_{\tilde a}$ 
into Eq.~(\ref{eq:InertialDampedHarmonicGauge}) would not be  what one might naturally think of as a
corotating damped harmonic gauge condition.   In order to clarify the meaning of the damped harmonic
gauge in a particular frame, we start by observing that the  quantity

\begin{eqnarray}
  \Delta_a \equiv \psi^{bc} \psi_{ad} \left( \Gamma^d{}_{bc} - {\Gamma^{(0)}}^d{}_{bc} \right) = \Gamma_a -  \psi^{bc} \psi_{ad}  {\Gamma^{(0)}}^d{}_{bc}
\end{eqnarray}
is a tensor, where ${\Gamma^{(0)}}^d_{bc}$ is a `background' connection associated with some background metric 
${\psi^{(0)}}_{ab}$.  
We can define the background in the inertial frame
to be the Minkowski metric, 
\begin{equation}
{\psi^{(0)}}_{ab} \equiv \eta_{ab} \; ,
\end{equation}
while the comoving background would be the tensor-transform of the inertial frame background metric
into the comoving frame,
\begin{equation}
{\tilde \psi^{(0)}}_{\tilde a\tilde b} =  { \psi^{(0)}}_{ a b} \; {J^a}_{\tilde a}  {J^b}_{\tilde b} ,
\end{equation}
with Jacobian
\begin{equation}
{J^a}_{\tilde b} = \frac{\partial x^a}{\partial \tilde x^{\tilde b}} .
\end{equation}
The transformation of $\Delta_a$ from the inertial to the corotating frame leads to
\begin{equation}
 \Gamma_a =  J_a{}^{\tilde a}{}  \tilde \Gamma_{\tilde a} - J_a{}^{\tilde a}{}  \tilde \psi^{\tilde b\tilde c}\tilde \psi_{\tilde a\tilde d}   {\tilde \Gamma^{(0)}}{}^{\tilde d}{}_{\tilde b\tilde c} .
\label{eq:GammaTransformationRule}
\end{equation}

The co-moving damped harmonic gauge condition is
\begin{equation}
  -\tilde \Gamma_{\tilde a} =  \mu_L \log\left(\frac{\sqrt{\tilde g}}{\tilde N} \right) \tilde t_{\tilde a} - \mu_S \tilde N^{-1} \tilde g_{\tilde a\tilde i} 
\tilde N^{\tilde i} .
\label{eq:GammaCorotDampedHarmonic}
\end{equation}
Substitution of Eq.~(\ref{eq:GammaCorotDampedHarmonic}) into Eq.~(\ref{eq:GammaTransformationRule}) and expressing the result in terms of
inertial frame metric quantities leads to
\begin{eqnarray}
  H_{a} =&&
  \mu_L \log\left(\frac{J \sqrt{g}}{N} \right) t_{a}
  - \mu_S N^{-1} g_{ai}
      \left( N^{i} - V^{i}\right)
      \\ && \nonumber
  + \psi^{bc} \psi_{ad}
    J_b{}^{\tilde b}
    J_c{}^{\tilde c}
    J^d{}_{\tilde d}
   \Gamma^{(0)}{}^{\tilde d}{}_{\tilde b\tilde c} ,
\end{eqnarray}
where $J=\det \left[ \frac{\partial x^{i}}{\partial x^{\tilde i}} \right]$ and $V^i = \frac{ \partial x^i }{ \partial{\tilde t} }$ is the coordinate
velocity of a comoving observer in the inertial frame.  When compared to the inertial-frame variant of the same gauge condition,
the $J$ term effectively makes the lapse condition milder early in the inspiral; the $V^i$ term damps the comoving frame shift to small values (rather than damping the inertial frame shift to small values regardless of the object's velocity);
and the last term is a consequence of the
transformation properties of $\Gamma_a$.

We have extensive experience with the inertial frame gauge condition.  Whether placing this in the comoving frame helps is yet to be explored.

\subsection{First Differential Order Form}

Historically there has been a lot more know-how available for the analytic and numeric treatment of first differential order systems.
Motivated largely by this reality, the main stream Generalized Harmonic evolution system in {\tt SpEC} has been
written in first order differential form\cite{Lindblom2006}. 
This is done by introducing the auxiliary variables $\Phi_{iab}$ and $\Pi_{ab}$ defined by
\begin{eqnarray}
N \Pi_{ab} &=& -  \partial_t \psi_{ab} + \gamma_1 N^i \Phi_{iab}  \\
\partial_i \psi_{ab} &=& \Phi_{iab} ,
\end{eqnarray}
where $\gamma_1$ is a free parameter.
The resulting evolution system takes the form
\begin{eqnarray}
\partial_t\psi_{ab}&-&(1+\gamma_1)N^k\partial_k\psi_{ab} 
 = - N\Pi_{ab}-\gamma_1N^i\Phi_{iab},
\label{e:psiEvol}\\
\partial_t\Pi_{ab} &-& N^k\partial_k\Pi_{ab} 
+ N g^{ki}\partial_k\Phi_{iab} - 
\gamma_1 \gamma_2 N^k \partial_k \psi_{ab}
\nonumber\\
&=&2N\psi^{cd}\bigl(  
  g^{ij} \Phi_{ica} \Phi_{jdb}
- \Pi_{ca} \Pi_{db}
- \psi^{ef}\Gamma_{ace}\Gamma_{bdf}
\bigr)
\nonumber\\&&
-2N\nabla_{(a}H_{b)}
- \frac{1}{2} Nt^c t^d \Pi_{cd}\Pi_{ab}
-N t^c \Pi_{c i} g^{ij}\Phi_{jab}\nonumber\\
&&+N\gamma_0 \bigl[2\delta^c{}_{(a}t{}_{b)}-\psi_{ab}
t^c\bigr] ({H}_c+\Gamma_c)
- \gamma_1 \gamma_2 N^i \Phi_{iab},\label{e:PiEvol}\\
\partial_t\Phi_{iab}&-&N^k\partial_k\Phi_{iab}
+N\partial_i\Pi_{ab}-N\gamma_2\partial_i\psi_{ab}
\nonumber\\
&=&\frac{1}{2} N t^c t^d \Phi_{icd}\Pi_{ab}
+Ng^{jk}t^c\Phi_{ijc}\Phi_{kab}
-N\gamma_2\Phi_{iab}\label{e:PhiEvol} ,
\end{eqnarray}
where $\gamma_0$ and $\gamma_2$ are additional free parameters of the system and
we have included all terms, including the nonprincipal part.

The advantage of this system is that it is a manifestly first order symmetric hyperbolic system\cite{Lindblom2006}. 
One of the drawbacks is that 
introduction of auxiliary variables leads to additional constraints for the system.  The new constraint quantities are
\begin{eqnarray}
\label{eq:Ciab}
{\cal C}_{iab} &=& \partial_i \psi_{ab} - \Phi_{iab}
\\
{\cal F}_a &\approx& t^c \partial_c {\cal C}_a = N^{-1} \left(-\partial_t {\cal C}_a - N^i \partial_i {\cal C})a\right)
\\
\label{eq:Cia}
{\cal C}_{ia} &\approx& \partial_i {\cal C}_a
\\
{\cal C}_{ijab} &=& 2\partial_{[i} \Phi_{j]ab} = 2 \partial_{[j} {\cal C}_{i]ab} ,
\end{eqnarray}
where full expressions for ${\cal F}_a$ and ${\cal C}_{ia}$ are given in Ref.~(\refcite{Lindblom2006}).
It has been shown\cite{Lindblom2006} that the constraints
associated with the extended, first order system form their own set of first order
symmetric hyperbolic equations with source terms
that vanish when the constraints are zero.  By implication, given vanishing initial and boundary data for the constraints, the constraint
propagation system will ensure that the constraints stay zero during the evolution.   In addition, we find  that, while
the unmodified constraint propagation system may be prone to exponentially growing modes (seeded by numerical error induced at each time-step), the constraint damping terms $\gamma_0,\gamma_1,\gamma_2$ are sufficient to control
these growing modes for all cases of interest.  Today, in all ``production-style'' binary black hole runs, we use the system given by
Eqs.~(\ref{e:psiEvol})-(\ref{e:PhiEvol}).

In parallel, as a side effort, a first differential order in time, second differential order in space version of the same system has been
implemented and was shown to work well for a limited number of test-cases\cite{Taylor:2010ki}.  
We expect that, if given enough effort, the second differential order form would be
be at least a factor of two more efficient, given the smaller number of evolution variables the code requires.
This approach has, however, received limited attention given
the success of the first order system.  

\subsection{Boundary Conditions}

One of the benefits of expressing a system in a symmetric hyperbolic form is that this formalism provides a way of identifying
the main characteristic speeds of a system across a given boundary, as well as the associated characteristic fields.  This then leads
to an easy-to-follow recipe for constructing boundary data.
One must provide data for those quantities that are incoming through the
boundary, while making sure that the outgoing quantities are updated through their evolution equations (and thus allowed to
naturally leave the domain).  An additional criteria is that the incoming data must not be specified in a way that injects constraint
violating modes.
There are a number of versions of constraint preserving boundary conditions for the generalized harmonic system. 
Most notably, Kreiss and 
Winicour has worked out a geometrically motivated, well-posed boundary system in Ref.~\refcite{Kreiss2006,Babiuc:2006ik,Kreiss:2013gw}. 

The {\tt SpEC} code implements a boundary algorithm based on Rinne, Lindblom and Scheel\cite{Lindblom2006}.
The current algorithm works well for BBH evolutions of up to a few dozen orbits. However, it becomes  a major limitation
in evolutions lasting hundreds of light-crossing times -- in these simulations our current boundary algorithm acts as
a gravitational potential, causing an acceleration of the center of mass.  The well-posedness and stability
of this evolution-boundary system has been studied by Rinne in Ref.~\refcite{Rinne:2006vv}.

\section{Domain Construction}

One key element influencing the design of the spectral domain used in the {\tt SpEC} binary black hole simulations
is the choice to use excision as a means of dealing with the space-time singularity inside the horizon.  Finite difference
codes have been successful in BBH evolutions with\cite{Pretorius2005a,Szilagyi2007}
and without excision\cite{Campanelli2006a,Baker2006a}.
While an excision-less domain is much easier to construct and an excision-less evolution algorithm is much less involved on
a technical level than one with excision, spectral accuracy would be lost if the evolution field had nonsmooth features.
The singularity would introduce such nonsmoothness, even if one factors out and regularizes the singularity (e.g., by evolving
the conformal metric and the inverse of the conformal factor, as done in some of the puncture codes).
As a result, in {\tt SpEC}
the computational domain needs to accommodate two inner spherical excision boundaries.   In addition, the most natural outer 
boundary shape is spherical as well, which is easily accommodated by the wave-zone spherical grid used in our code.

\begin{figure}
\centerline{\includegraphics[scale=0.27]{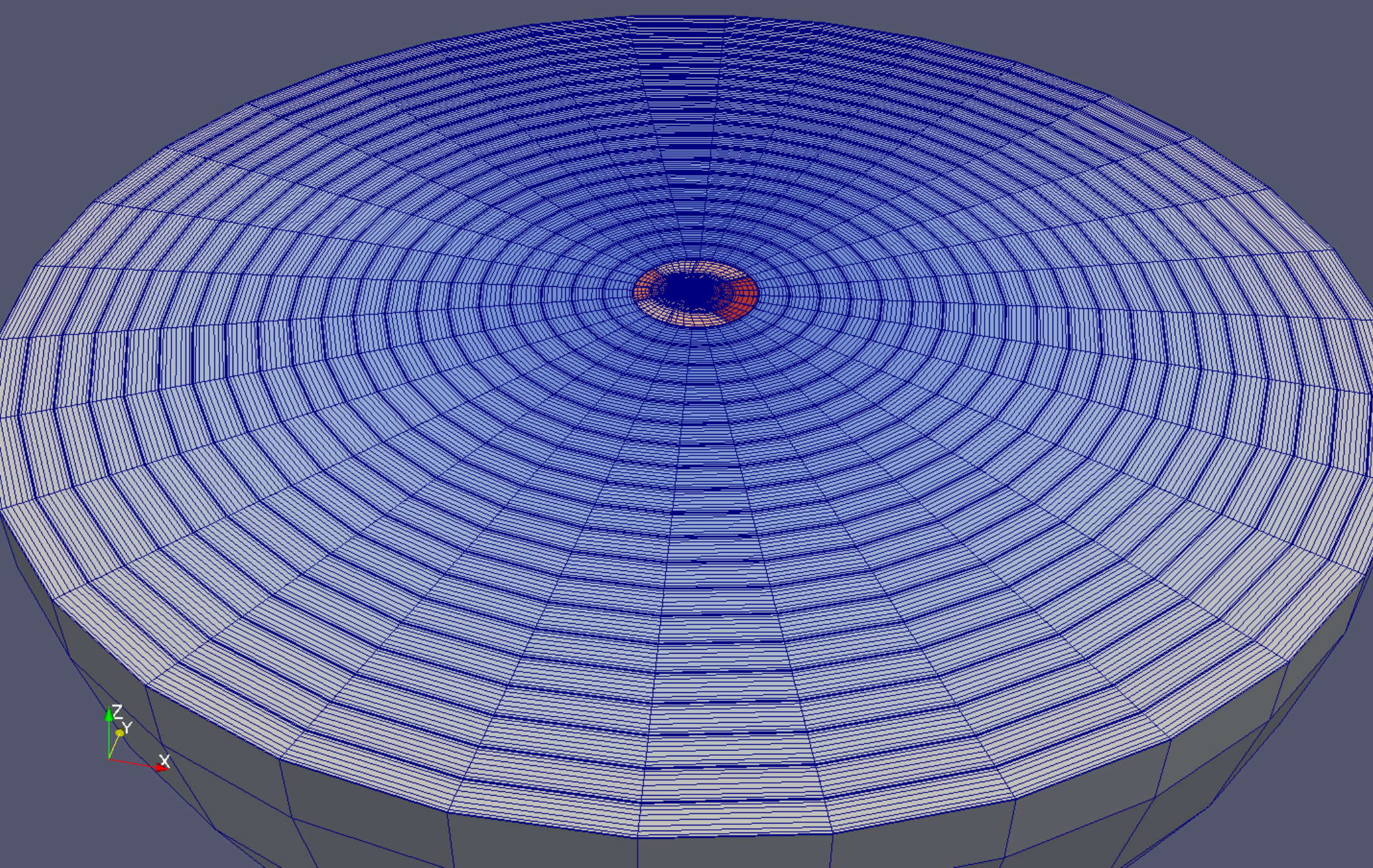}
\includegraphics[scale=0.3075]{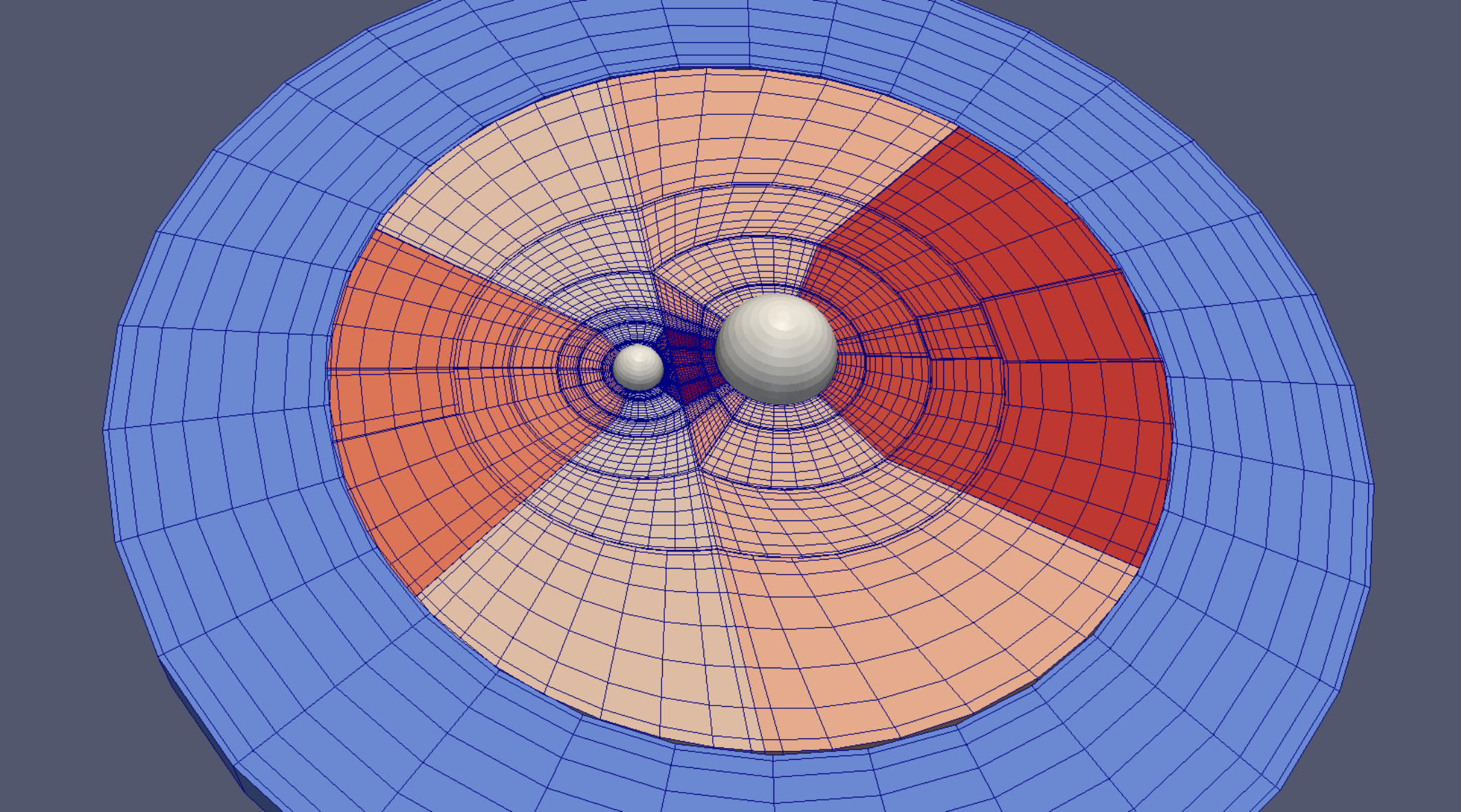}
}
\caption{\label{fig:CutSphereFullHalfGrid} 
{ Left:} the $z\leq0$ portion of our spectral gird structure used for BBH runs.
The outer region consists of a set of spherical shells centered at the coordinate origin.  The angular spectral
representation in these shells consists  of a scalar spherical harmonic expansion, while radially we use 
Chebyshev Gauss Lobatto collocation points. These spherical shell subdomains are labeled as {\tt SphereC0, SphereC1,...}
in our code.
{ Right:}
A closer look at the interior of the
grid used in our BBH runs.  This plot displays the $z\leq0$ portion of the innermost wave-zone shell, {\tt SphereC0}, 
 surrounding a set of distorted cylinders and two inner spherical grid-structures, centered around the individual
black-holes.  The 
axes of the cylinders are aligned with the coordinate $x$ axis. These shapes are distorted such that the lower disk-shaped
boundary of the cylinder is touching interior spherical shells centered around the individual black holes, while
their outer boundary is touching the inner spherical boundary of {\tt SphereC0}.  The three-dimensional figure can be obtained
by a $180^{\circ}$ rotation around the $x$ axis. }
\end{figure}

\begin{figure}
\centerline{
\includegraphics[scale=0.3]{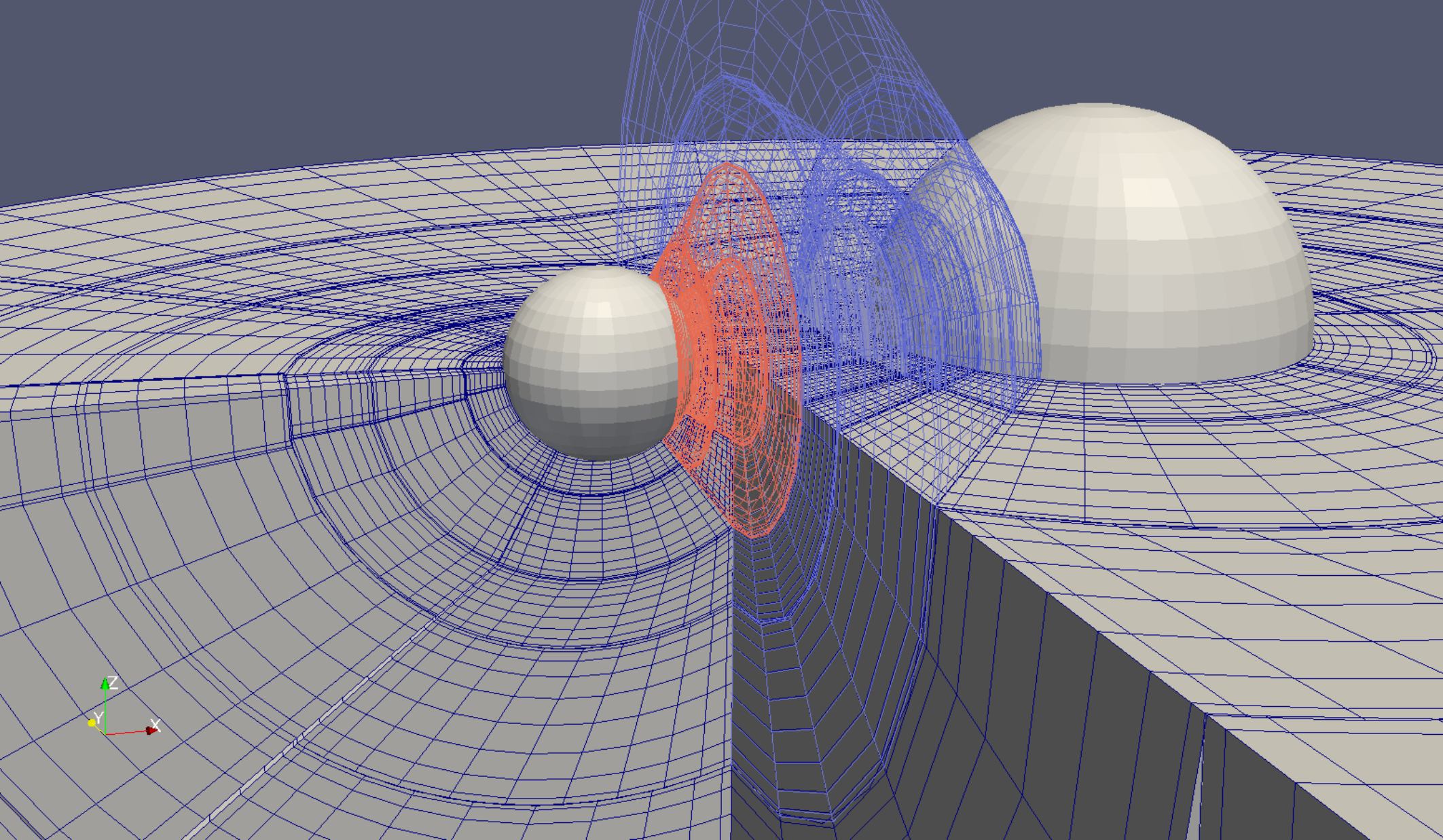}
\includegraphics[scale=0.276]{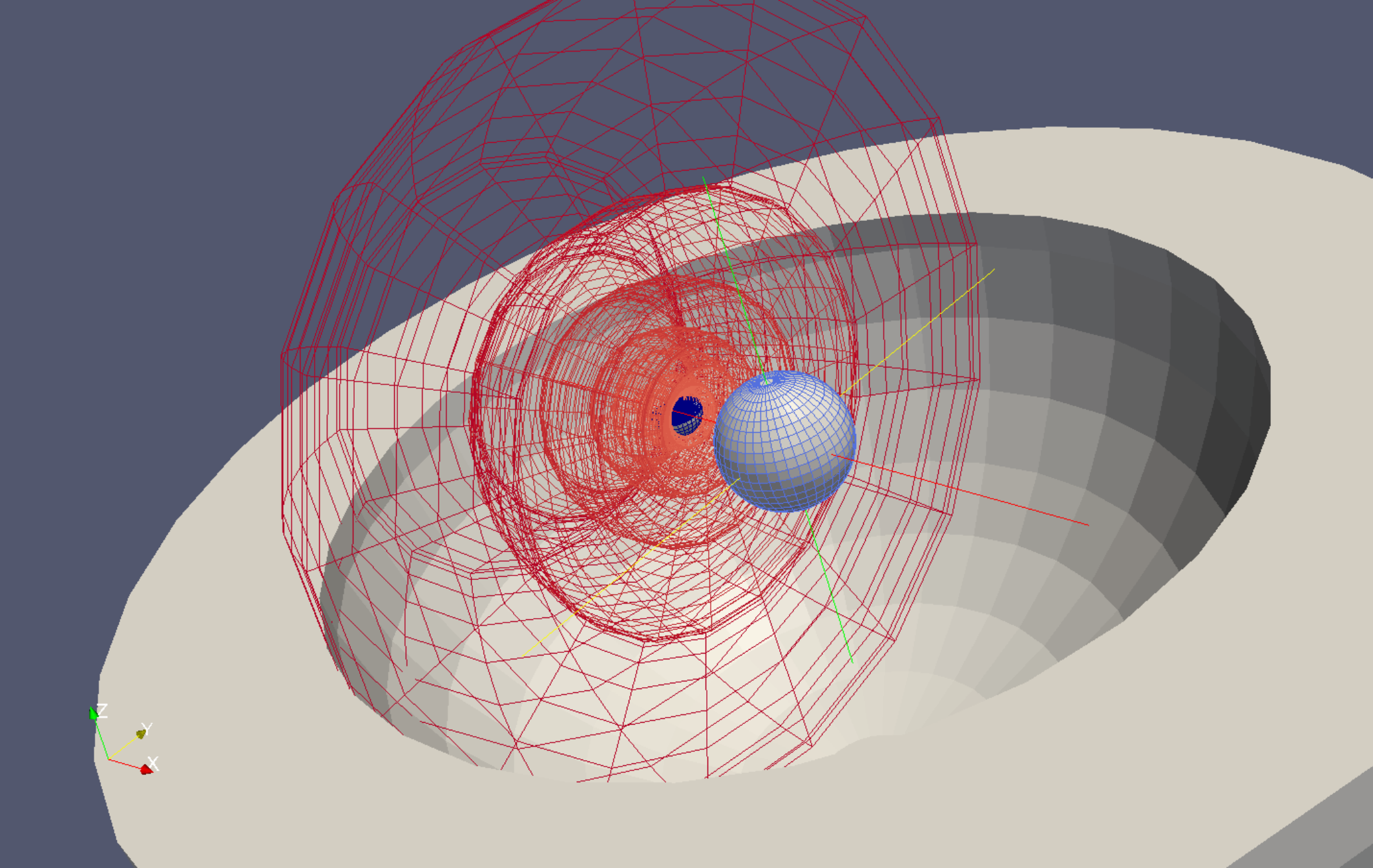}
}
\caption{\label{fig:CutSphereCylindersM} (Color online)
{ Left:} This plot reveals details of the grid in the immediate neighborhood
of the two sets of spherical shells, centered  around the individual black holes (shown as gray spheres).
The view reveals an inner set of cylinders that touch, on one end, the outermost of these inner sets
of shells, and on the other end an $x=$const. grid-plane, referred to as the {\tt CutX} plane in our code. 
{ Right:} In this view we have removed the cylinders associated with 
the larger black hole and are providing a $3D$ grid-frame view of the cylinders around the smaller black hole. The close end
of the cylinders shown here are all touching  the {\tt CutX} plane. The far end of the cylindrical structure
touches the inner boundary of {\tt SphereC0}, displayed in grey.}
\end{figure}

\begin{figure}
\centerline{
\includegraphics[scale=0.29]{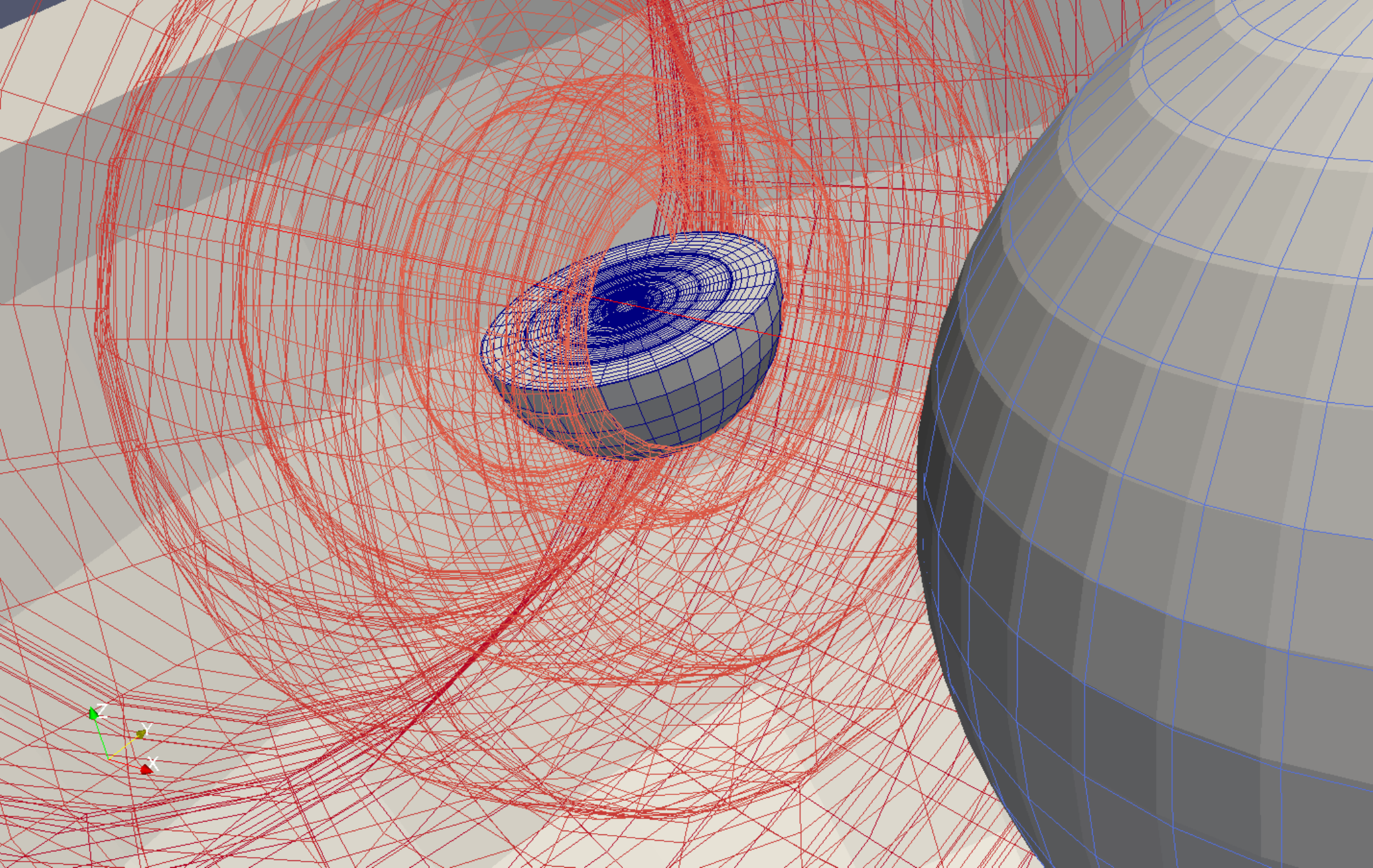}
\includegraphics[scale=0.29]{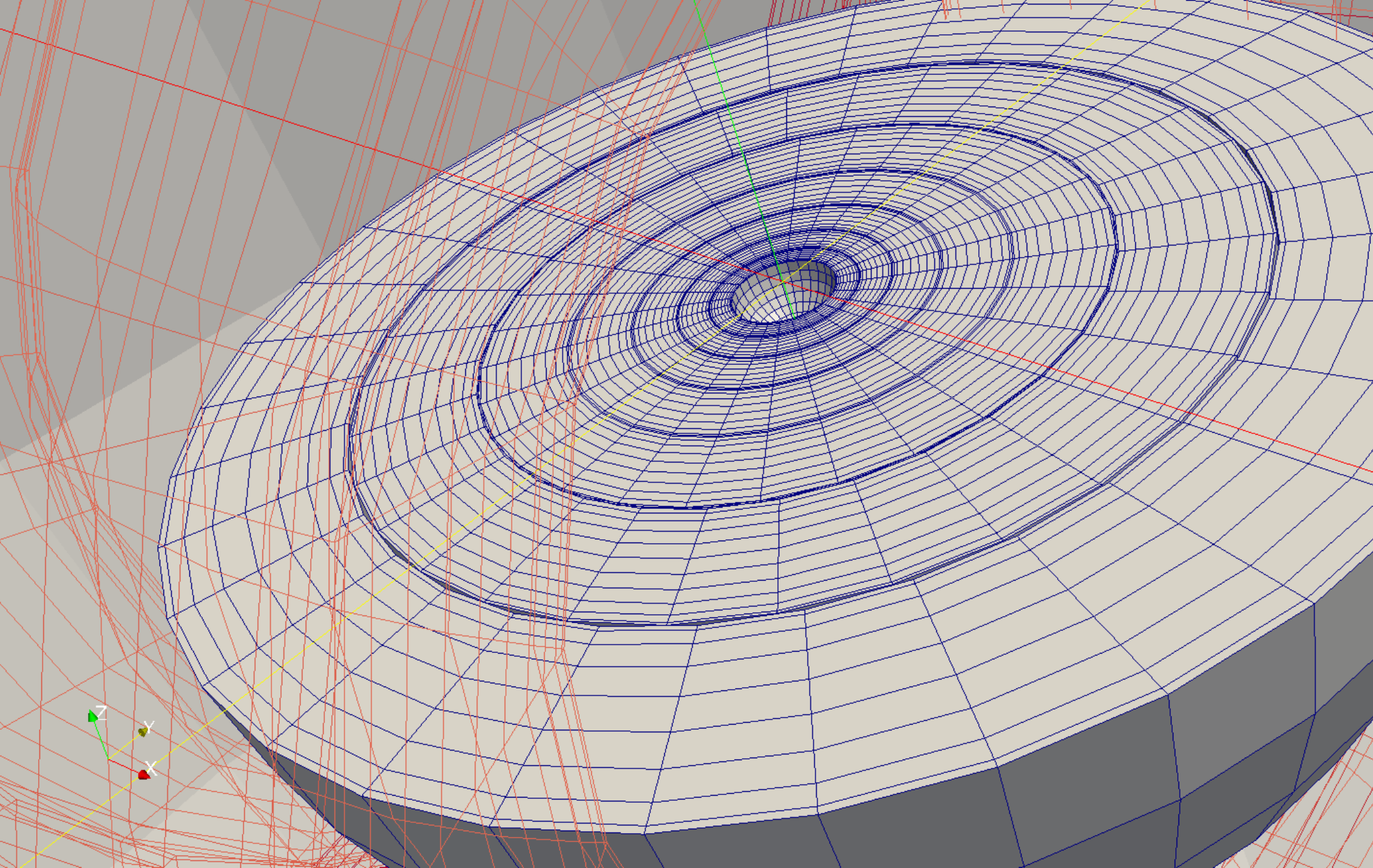}
}
\caption{\label{fig:CutSphereCylindersAroundSphereB} (Color online)
{ Left:} This figure illustrates the $z<0$ portion of the
spherical shells centered around the smaller black hole as well as the cylinders surrounding it (shown as
a red wire frame).  The large spherical surface on the close right side is the outer boundary of the shell structure
around the larger black hole. 
The inner sets of
spherical shells, around the individual black holes, are labeled {\tt SphereA0, SphereA1...} and {\tt SphereB0, SphereB1,..}.
Both of these sets of shells are concentric, with their centers near the $x$ axis. (The exact location of the black holes
is determined by
the elliptic initial-data solver, as this determines the location of the black holes at $t=0$, and the shells are
concentric with the associated black hole apparent horizon.)
{ Right:} As a last illustration of our BBH grid, we show, from a close view,
the excision boundary within the {\tt SphereB0} subdomain (which is the innermost of the shells around the smaller black hole).}
\end{figure}

After a number of attempts at the problem of domain construction, 
we have settled on the non-overlapping domain described in
the Appendix of
Ref.~\refcite{Buchman:2012dw}. This grid is composed of a small number of
spherical shells around each excision boundary,
labeled as {\tt SphereA}$n$ and {\tt SphereB}$n$, where $n=0,1,\ldots$
 (see Fig.~\ref{fig:CutSphereCylindersAroundSphereB}).  The innermost of these sets of
 spherical shells, also referred to as the excision
subdomain, has index zero.  In addition, the grid contains
a third set of shells describing the gravitational wave zone  (shown in Fig.~\ref{fig:CutSphereFullHalfGrid}),
labeled as {\tt SphereC}$n$, $n=0,1,\ldots$.  In a typical simulation we have $20$ outer shells.
The space between the innermost outer shell, {\tt SphereC0} and the outer boundary of the inner shell structures
is filled up by a set of cylindrical subdomains, distorted both on the lower and upper end, so that they touch
their neighboring subdomain  (shown in 
Fig.~\ref{fig:CutSphereFullHalfGrid},
and
Fig.~\ref{fig:CutSphereCylindersM}).
The space between the two inner sets of spherical shells is filled in with another set of distorted cylinders
which, on one end, are touching the spherical boundary of the outermost of {\tt SphereA} and {\tt SphereB},
while on the other end they touch an $x=$const grid-plane, referred to as the {\tt CutX} plane, 
as shown on Fig.~\ref{fig:CutSphereCylindersM}.
 
In conjunction with inter-block
penalty boundary conditions and appropriate spectral filters (see Sec.~\ref{sec:SpectralNum}),
a major advantage of our current compact binary grid is its robust stability
with respect to high frequency noise generated by our numerical update scheme.
This is a non-trivial property of a spectrally
evolved BBH system where high accuracy leaves room for very little numerical dissipation.
In addition, we find our domain to adapt well to binaries with very different masses.\cite{Mroue:2013xna,Buchman:2012dw}

The domain construction (or control of its parameters) requires an extra amount of care near the merger
of the binaries.   This is to be expected, as BBH simulations tend to spend the longest amount of physical and
wall-clock time in the inspiral phase. During this phase, little changes as the black holes slowly approach each other,
while radiating energy via gravitational wave emission.  It makes most sense to design a grid that does well during the early
stage.
As the binary approaches merger, the excision boundaries (kept in the near vicinity of the individual apparent horizons) approach
each other.  It is essential that all features of the grid are able to adjust to this deformation.   In order to avoid grid singularities
(or noninvertible maps, as described in Sec.~\ref{seq:ControlSystem}), at various stages during
the plunge our algorithm defines
a new grid, with modified parameters  such that the position of the {\tt CutX} plane remains
between the spherical shells around the excision boundaries until after merger.  
See Secs.~\ref{sec:rTypeAmr} and \ref{sec:ShellDrop} for details.

\section{Control System}
\label{seq:ControlSystem}

As mentioned earlier, the {\tt SpEC} code uses excision for evolving systems involving black holes.   
This creates its own set of challenges.  In a finite difference code, in order to properly excise the interior
of a black hole, one needs to locate the apparent horizon and implement an algorithm with a
moving excision boundary.  As the black hole moves on the grid, there will be grid-points that
are evolution points on one time-step which become excision points on the next time level. In addition, there will be
other points that are labeled as excision points at some time only to become evolution points as the black hole
moves past them.   This can be implemented on a point-wise basis, as has been done in a number of finite difference codes,
e.g. in  Ref.~\refcite{Szilagyi2007}.  In a spectral algorithm the removal of a single grid-point is not feasible as, in some sense,
spectral differencing stencils are global within the subdomain (or block).  As a result, there  is no option to move an
excision boundary along a particular direction by one grid-point.  Rather, we employ a feed back
control system\cite{Hemberger:2012jz}
based on measurement of the position of the black holes, which
rotates, translates and shrinks the grid such that the coordinate
centers of the excision boundaries tightly follow the motion of the coordinate centers of the apparent horizons.  In addition,
we monitor the coordinate shape of the apparent horizons in the co-moving frame and apply a distortion map to the 
excision boundary such that it stays just inside the apparent horizon throughout the
evolution, including plunge and merger.
The full sequence of maps employed in a BBH evolution is given by\cite{Hemberger:2012jz}
$x^k=\mathcal{M} (\hat x^{\hat k})$, where $\{x^k\}$ are inertial coordinates, $\{\hat x^{\hat k}\}$ are grid-frame coordinates, and 
\begin{equation}
  \label{eq:MapSequence}
\begin{array}{ll}
{\mathcal M}_{{\rm Grid}\rightarrow{\rm Inertial}} =&
\map{Translation}
\circ\map{Rotation}
\circ\map{Scaling}  \\
& \circ\;\map{Skew}
\circ\map{CutX}
\circ\map{Shape}.
\end{array}
\end{equation}
Here
\begin{itemize}
  \item
    $\map{Translation}$ is a spatial map which
    translates the mass-weighted average
    of the centers of the excision boundaries
    so that this point is in the neighborhood of the coordinate-center of mass
    of the BBH system as it evolves. This map leaves the position of the outer boundary unchanged.
  \item
    $\map{Rotation}$ rotates the grid such that the line connecting
    the centers of the excision boundaries is aligned with the one connecting the 
    coordinate centers of the apparent horizons.
  \item
    $\map{Scaling}$ shrinks (or expands) the grid at the right rate such
    that the distance between the centers of the excision boundaries is synchronized
    with the distance between the coordinate centers of the apparent horizons.
    This map is also responsible for a slow inward motion of the outer boundary
    so that constraint violating modes with vanishing characteristic speed
    leave the domain rather than stay forever.
  \item
    \map{Skew} is responsible for distorting the {\tt CutX} plane at late times, when
    the excision boundaries are near each other, such that this skewed grid plane
    stays at a finite distance from the excision boundaries even when they
    are non-perpendicular to the $x$ axis. See Figure~3 in Ref.~\refcite{Hemberger:2012jz}
  \item
    \map{CutX} is responsible for controlling the position of the {\tt CutX}
    plane (by moving along the $x$ axis) such that it remains at a
    finite distance from the point where the excision boundaries
    cross the $x$ axis. See Sec.~\ref{sec:rTypeAmr} as well as 
    Ref.~\refcite{Hemberger:2012jz} for details.
  \item
    \map{Shape} is responsible for deforming the excision boundary from its
    original spherical shape to that of the apparent horizon.
\end{itemize} 

In  Sec.~\ref{sec:DampedHarmonicGauge} of the paper we made reference to the
 {\em comoving} frame (with coordinates $\{\tilde x^{\tilde k}\}$)
 defined as the frame in which the coordinate centers of the apparent
horizons are at a fixed position.  This frame is connected to the inertial frame by 
the map
\begin{eqnarray}
{\mathcal M}_{{\rm Comoving}\rightarrow{\rm Inertial}} &=& \map{Translation} \circ \map{Rotation} \circ \map{Scaling} \; ,
\label{eq:ComovingFrame}
\end{eqnarray}
leading to
\begin{eqnarray}
x^k_{\rm Inertial} &=& 
 \map{Translation} \left( \map{Rotation} \left( \map{Scaling} \left(\tilde x^{\tilde k}_{\rm Comoving}\right)\right)\right)
\; .
\end{eqnarray}
Another frame of relevance is the {\em distorted} frame $\bar x^{\bar k}_{\rm Distorted}$, connected to the
grid-frame coordinates $\hat x^{\hat k}_{\rm Grid}$ by
\begin{equation}
\label{eq:DistortedFrame}
{\mathcal M}_{{\rm Grid}\rightarrow{\rm Distorted}} =   \map{CutX} \circ \map{Shape}\; ,
\end{equation}
giving
\begin{equation}
\bar x^{\bar k}_{\rm Distorted}=   \map{CutX} \left( \map{Shape}  \left( \hat x^{\hat k}_{\rm Grid}\right)\right) \; .
\end{equation}
This is the frame in which we search for the apparent horizons, feeding it into the control system responsible for
updating the  time-dependent parameters of the various maps.
In terms of the maps associated with the distorted and the comoving frames the full transformation from the grid frame to
the inertial frame  can then be written as
\begin{eqnarray}
x^k_{\rm Inertial} &=&  {\mathcal M}_{{\rm Grid}\rightarrow{\rm Inertial}} \left( \hat x^{\hat k}_{\rm Grid} \right)
\\
                                   &=&   
 {\mathcal M}_{{\rm Comoving}\rightarrow{\rm Inertial}} \left(
 {\mathcal M}_{{\rm Skew}} \left(
 {\mathcal M}_{{\rm Grid}\rightarrow{\rm Distorted}} \left(
 \hat x^{\hat k}_{\rm Grid} 
\right)
\right)
\right) .
\end{eqnarray}

Note that for the inspiral part of our BBH simulations, where 
$\map{CutX}$ is inactive, the distorted frame is simply the grid frame distorted under the action
of the shape-map $\map{Shape}$.  Also, note that the comoving frame and the distorted frame differ
by the skew-map $\map{Skew}$. The motivation of working with two different though similar
frames is in their use.
The comoving frame is instrumental in defining a gauge condition, as described in
Sec.~\ref{sec:DampedHarmonicGauge}.
The distorted frame is introduced for the sake of a more convenient implementation of our grid-control
algorithm; thus its definition is more naturally stated through its relationship to the grid frame.

\section{Pseudo-Spectral Numerical Algorithm}
\label{sec:SpectralNum}

The irreducible topologies at the core of our grid construction are 
\begin{eqnarray}
I_1 &=& \left\{ x\in {\mathbb R}| a\leq x\leq b\right\} , \\
S_1&=&\left\{(x,y)\in {\mathbb R}^2|  x^2+y^2=a^2  \right\},\\ 
S_2&=&\left\{(x,y,z)\in {\mathbb R}^3|  x^2+y^2+z^2=a^2  \right\},\\
B_2&=&\left\{(x,y)\in {\mathbb R}^2|  x^2+y^2\leq a^2  \right\} . 
\end{eqnarray}
The computational domain for binary black hole simulations is built of blocks that are topologically 
\begin{equation}
I_1 \times S_2,\;  I_1 \times S_1 \times I_1,   \quad \mbox{and} \quad I_1 \times B_2.  
\label{eq:TopProduct}
\end{equation}
The spectral basis is associated
with the irreducible topologies $I_1, S_1, S_2, B_2$  is, in order, Chebyshev polynomials,
Fourier expansion, scalar spherical harmonics and one-sided Jacobi polynomials.  Each of these
come with their set of collocation points so as to optimize the conversion between collocation point data
and the spectral coefficient representation.    Our algorithm stores collocation values of each evolved quantity,
at a given time-step. 

Derivatives are computed by forming the spectral coefficient representation of the same
data and then recombining those with the analytic derivatives of the spectral basis functions.   
For example, given some smooth function $f(x)$ we approximate it as an expansion in terms of a spectral basis $b_i(x)$,
\begin{equation}
f(x) \approx \sum_i c_i b_i(x) \, .
\end{equation}
The derivative of the function will then be approximated by
\begin{equation}
\frac{\partial}{\partial x}f(x) \approx \sum_i c_i  \frac{\partial}{\partial x}  b_i(x)
\end{equation}
where the derivatives of $b_i(x)$ are known analytically.

Given the nonlinear
nature of the evolution system of interest, we find it easier to 
compute the RHS of the equations, at each collocation point,
using the actual function values (rather than their spectral coefficients).   The nonlinearity implies coupling between
neighboring coefficients.   By implication, the highest coefficients of any given representation will effectively be updated
by the ``wrong equation.''  In other words, their effective update scheme depends on whether there exists a higher coefficient
in the data or not.   In some cases this is not a problem since, for well-resolved, exponentially convergent representations of a 
smooth evolved field, the highest coefficient may already have a very small magnitude so that
having (or not having) a next higher
coefficient would have no appreciable impact on the evolution of the field.  In order to achieve robust stability,
we find that, we must
filter these higher coefficients.  In particular, after each time-step we set the highest four coefficients of the $Y_{lm}$ expansions
to zero\cite{Scheel2006}, 
\begin{equation}
c_{lm} \rightarrow c_{lm} \times  H[L_{\max}-l-4]  ,
\label{eq:HeavisideFilter}
\end{equation}
where $H[n]$ is the Heaviside step-function.  
In addition, we use the exponential filter
\begin{equation}
c_k \rightarrow c_k \exp\left[-\alpha \left(\frac{k}{N-1}\right)^{2p}  \right], \quad k=0, \ldots, N-1
\label{eq:ExpChebFilter}
\end{equation}
with $\alpha=36,p=32$ for the Chebyshev coefficients, as suggested in Refs.~\refcite{Gottlieb2001}
and \refcite{Hesthaven2007}. 
We use the same exponential filter for the 
Fourier coefficients as well, using $\alpha=36,p=24$, 
though in this case our choice is entirely heuristic.
These filters are applied to the evolution variables $\psi_{ab}, \Pi_{ab}, \Phi_{iab}$ after each full time-step.
In addition,  
each time one subdomain provides boundary data to another via interpolation,
this interpolated data is filtered using the Heaviside filter Eq.~(\ref{eq:HeavisideFilter}) for the scalar spherical
harmonic expansions on $S_2$ type boundaries and 
the exponential
filter Eq.~(\ref{eq:ExpChebFilter}) for all Chebyshev and Fourier expansions. For those boundaries, 
where the neighboring collocation grids are aligned at the boundary, there is no need for interpolation.
In these cases we are copying the data from one subdomain to the other without any filtering.
Having the right filtering scheme proved crucial in being able to establish robust stability.

\subsection{Accuracy Diagnostics}

\begin{figure}[h]
\centerline{\includegraphics[scale=0.24]{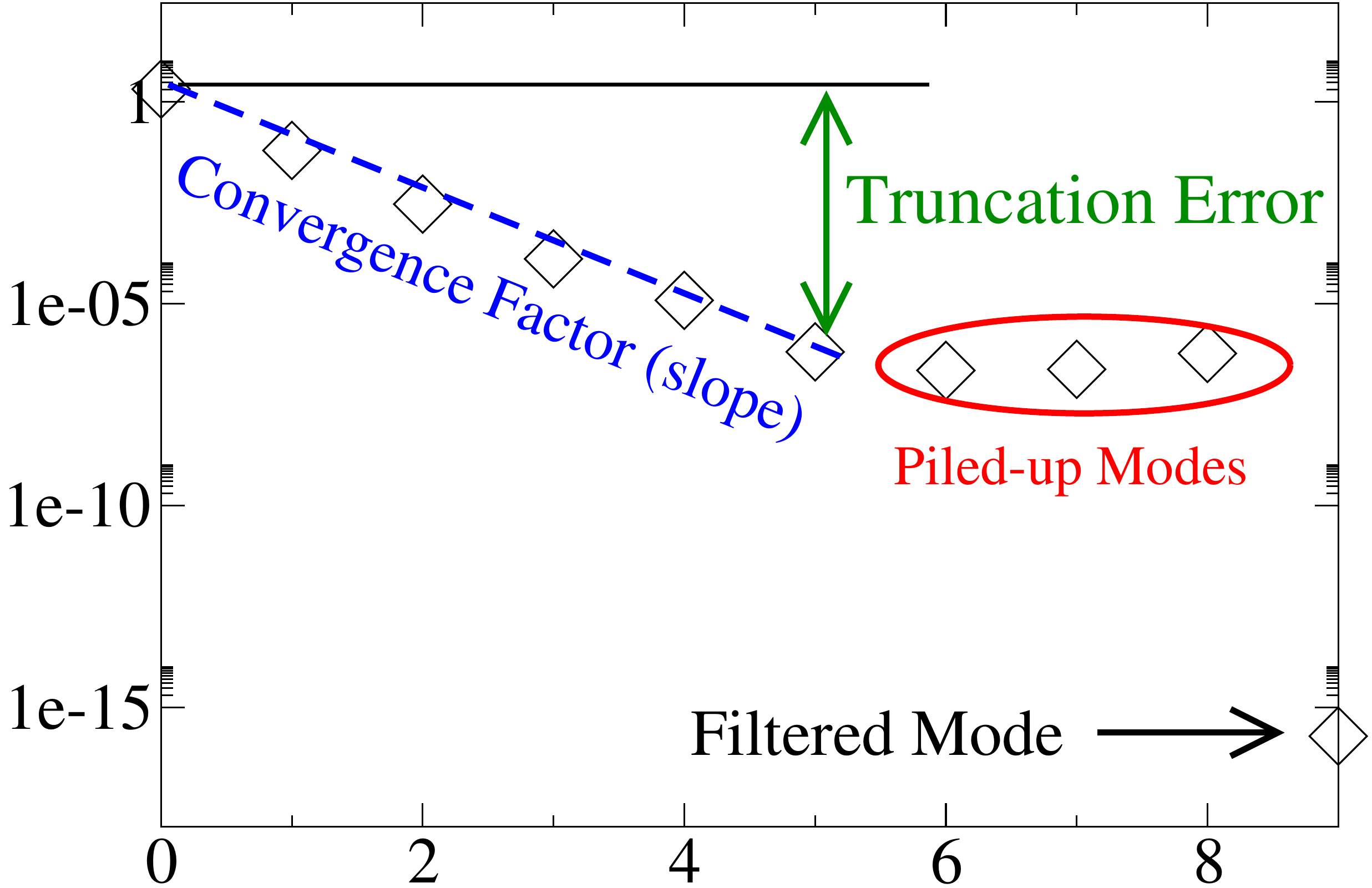}}
\caption{\label{fig:PowerMonitor}Plot of a typical power-monitor. The first six points (counting from left to right)
show exponential convergence of the spectral expansion. The slope (on the linear-log plot)
of these points defines 
of the {\em convergence factor} of the spectral expansion. 
The next three points show no convergence. They are designated
as {\em  pile-up modes}. 
The point on the far right, with $O($roundoff$)$ value, is a coefficient that is reset by a filtering algorithm at the end of
each time-step. We call this a {\em filtered mode}.}
\end{figure}

Exponential convergence of the spectral representation of smooth data in a particular basis requires that, for sufficiently large
mode number, the 
expansion coefficients are exponentially decaying functions of the mode number.
As an example, for a smooth function on an $S_2$ surface,
the spectral expansion in terms of scalar spherical harmonics is
\begin{equation}
  f(\theta,\phi) = \sum_{l=0}^{L} \sum_{m=-l}^{l} C_{lm} Y_{lm}(\theta,\phi) + \mbox{residual} .
\end{equation}
When the function $f$ is well resolved, then the magnitude of the
highest order coefficients will decay exponentially,
\begin{equation}
   C_{l,m} \sim e^{-l}, \quad \mbox{for large} \; l  .
\end{equation}
This suggests that one should monitor the quantity
\begin{equation}
\label{eq:YlmPowerMonitor}
P_{l} = \sqrt{\frac{1}{2m+1} \sum_{m=-l}^l | C_{l,m} |^2}
\end{equation}
in order to asses the accuracy of the spectral representation.
This is done in a more generic context. Whenever dealing with tensor products of irreducible topologies, we
monitor accuracy of the representation by computing the average (in an $L_2$ sense)
over all other dimensions; e.g., for $I_1 \times S_1 \times I_1$, we define power-monitors according to the
three indexes, $P_{k_0}, P_{k_1}, P_{k_2}$ as
\begin{equation}
\label{eq:GenericPowerMonitor}
P_{k_0} = \sqrt{ \frac{1}{N_1 N_2}   \sum_{k_1,k_2} \left|C_{k_1,k_2,k_3} \right|^2} ,
\end{equation}
with similar definitions for $P_{k_1},P_{k_2}$.
A time-snapshot of an example power-monitor plot is provided in Fig.~\ref{fig:PowerMonitor}.
A few key features are immediately noticeable.  The highest mode has $O($roundoff$)$ power,
which results from filtering.   
The lowest six modes (or: the first six, counting from the left)  in Fig.~\ref{fig:PowerMonitor}
show a clean exponential decay.  This is indicative of exponential convergence, and allows us
to define the convergence factor as the logarithmic slope of the power in the coefficient vs. its index.
The ratio of the highest to the lowest coefficient in this clean convergent set of points gives a measure of the 
truncation error in the expansion.   And the few extra modes that are placed between the convergent modes and the filtered
mode are labeled as ``pile-up modes.''  These indicate saturation of the grid with respect to the data that is
being spectrally expanded.  They are most likely related to noise in the data, possibly due to non-smooth boundary data
on the subdomain boundary. 

\begin{figure}
\centerline{\includegraphics[scale=0.4]{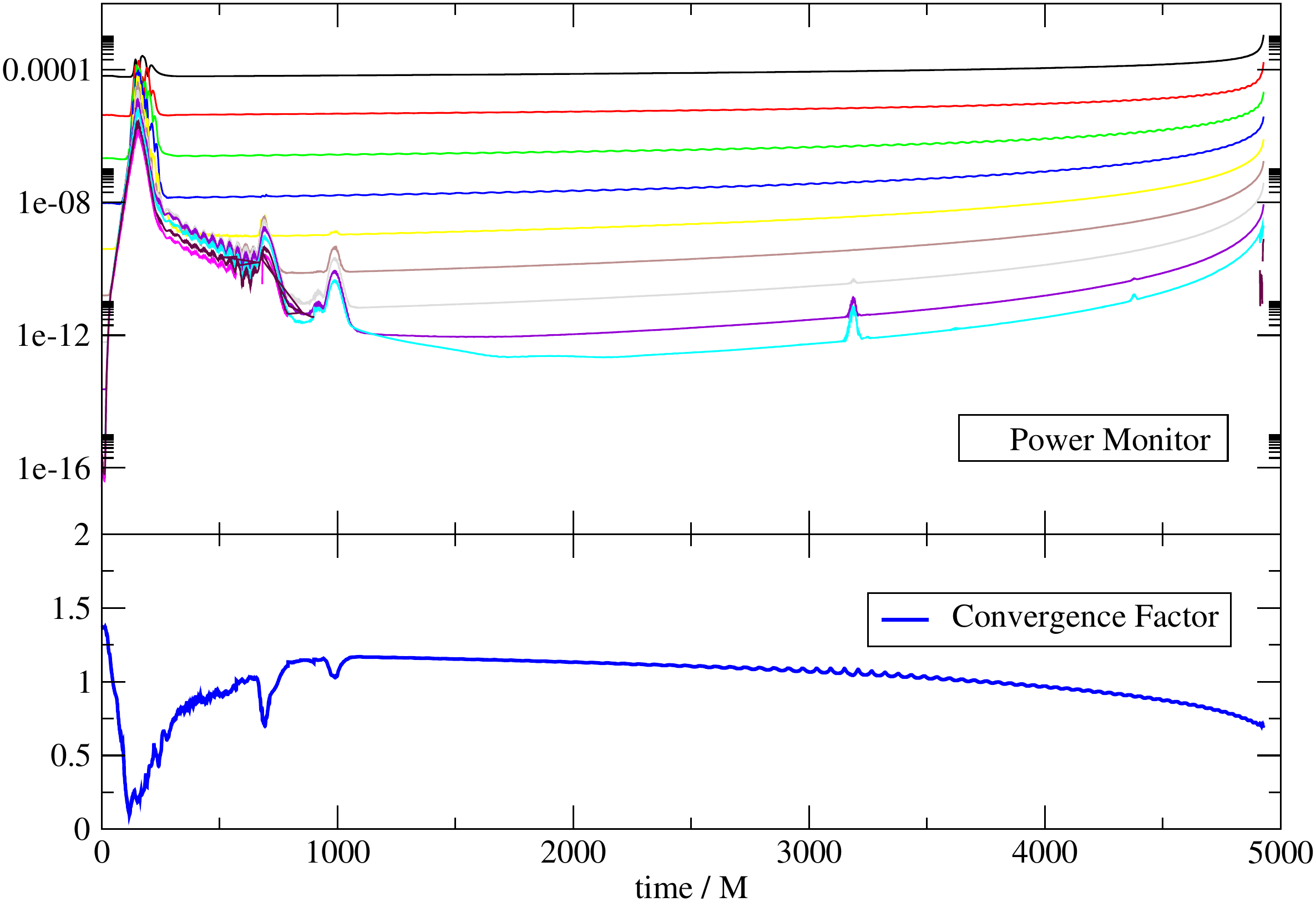}}
\caption{\label{fig:ConvFactor} Convergence factor for the radial spectral expansion of the {\tt SphereC5} wave-zone
spherical shell for a non-spinning $22.5$ orbit BBH run, with mass ratio $q=7.1875$.
The upper panel shows the power-monitor  Eq.~(\ref{eq:GenericPowerMonitor})  associated with the radial  $I_1$ expansion.  
The noise in the early part of the simulation is caused by the
junk radiation (high frequency radiation seen in the early part of BBH evolutions, caused by unphysical content of the initial data)
passing through the grid. The lower panel shows the convergence factor Eq.~(\ref{eq:ConvFactor}) for this
same power-monitor.   As the junk radiation passes through the grid, the convergence factor drops by nearly one order
of magnitude.   However, as this high frequency wave leaves the grid, the convergence factor rebounds and
stays $O(1)$ for the rest of the evolution.  
}
\end{figure}

For spectral Adaptive Mesh Refinement (AMR), it is important to define all of these quantities as smooth functions
of the power monitor values $P_k,  k=0,\ldots,N_k-1$.   Let ${\cal S}[k_1,k_2]$ be the slope of  the least square fit of 
$\log_{10}(P_k) = f(k)$ 
for the points $k=k_1\ldots k_2$, with $0\leq k_1<k_2\leq N_k-1$.
Let ${\cal E}[k_1,k_2]$ be the error in this fit. 
In our current algorithm we first compute ${\cal S}[k_1,k_2]$ and ${\cal E}[k_1,k_2]$
for a variety of selections of data points.  Then we define the 
convergence factor (up to an overall negative sign) 
as the average of the slopes, weighted by the inverse of their fit error,
\begin{equation}
\label{eq:ConvFactor}
{\cal C}[P_k] \equiv -
\left.
\sum_{
   \begin{subarray}{c}
         k_1 = 0, 2 \\
         k_2 = k_1+4 ,  \tilde N_k -1
   \end{subarray}
                              }
{ \frac{ {\cal S}[k_1,k_2]   } { \epsilon+ {\cal E}[k_1,k_2]    } } 
\right/
\sum_{
   \begin{subarray}{c}
         k_1 = 0, 2 \\
         k_2 = k_1+4,  \tilde N_k -1
   \end{subarray}
                              }
{ \left( \epsilon+ {\cal E}[k_1,k_2]    \right)^{-1} } 
,
\end{equation}
where $\epsilon$ is a small positive constant to avoid division by zero in case the linear fit has no error, while $\tilde N_k$ 
is the number of unfiltered modes, with $\tilde N_k \leq N_k$.   Note that for some of the filters, in particular for the 
scalar $Y_{lm}$ filter used in the evolution of tensor fields,  the filtering algorithm amounts to setting the highest few coefficients
to zero, as shown in Eq.~(\ref{eq:HeavisideFilter}).  In this case the number of filtered modes $N_k - \tilde N_k$
would equal the number of those coefficients that are being reset.   In the case of
the exponential Chebyshev filter Eq.~(\ref{eq:ExpChebFilter}) we approximate the number of filtered modes as those whose
coefficient is below some threshold, e.g., it is $O($roundoff$)$.  
By calculating a large number of fits $S[k_1,k_2]$ and weighting them by the inverse of the accuracy of the fit,
we give the larger weight to those sets of points that provide a clean slope.
This is important on both ends of the spectrum.  For the lower order modes one cannot expect, generically,
a particular behavior (such as exponential decay) as we look at the spectral expansion of an arbitrary (smooth)
function.\footnote{   
For instance, the space-time metric in the weak field region (such as the neighborhood of the outer boundary
in a typical binary black hole simulation) has the form
\begin{equation}
\psi_{ab} = \eta_{ab} \quad +\quad \mbox{small perturbation}
\end{equation}
(where $\eta_{ab}$ is the Minkowski metric).
In this case the lowest order coefficient corresponding to the constant element of the expansion basis will be
$O(1)$ while all higher coefficients will be several orders of mangitude smaller. 
Resolving the perturbative part of the metric in this wave zone is essential, as this is the effect we are trying to capture and
translate into the signal seen by a gravitational wave detector. 
The convergence factor estimate would be incorrect if it included the
$O(1)$ part coming from the flat space metric $\eta_{ab}$.}
The highest modes may be tainted by noise (they could be pile-up modes) which, again, would lead
to inaccurate measure of the convergence factor.  
This algorithm also gives
some level of robustness when certain coefficients show anomalies (e.g. because the data has some excess power in a given
spectral mode).  An example power-monitor (as a function of time), along with the associated convergence factor computed
using Eq~(\ref{eq:ConvFactor}), is given in Fig.~\ref{fig:ConvFactor}.

\begin{figure}
\centerline{\includegraphics[scale=0.4]{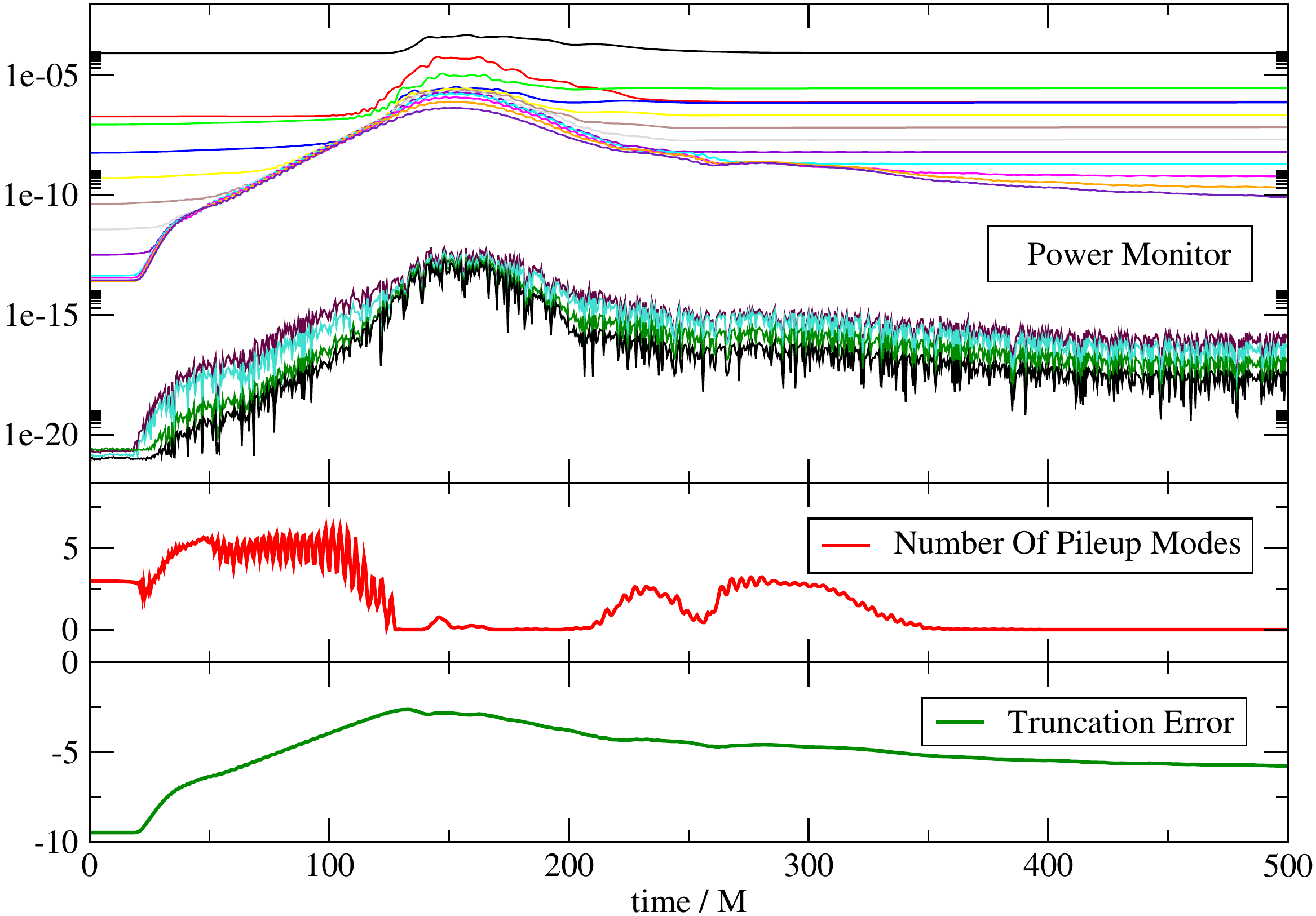}}
\caption{\label{fig:TruncErrAndPileUp}
 The {top} panel shows 
the power-monitor  Eq.~(\ref{eq:GenericPowerMonitor})  associated with the angular $S_2$ grid 
of the {\tt SphereC5} wave-zone spherical shell 
for the first $500M$ evolution of a non-spinning $22.5$ orbit BBH run, with mass ratio $q=7.1875$.
 The four noisy modes with $O($roundoff$)$ values are the filtered
coefficients.  As the junk radiation passes through, the unfiltered
modes pile-up on each other. Eventually the junk radiation leaves the grid, however, and convergence of the higher modes
is recovered.  {Middle}
panel:  the number of pile-up modes,  Eq.~(\ref{eq:PileUpMode}), show very clearly when do the power spectrum coefficients 
clutter next to each other.   {Bottom} panel: the truncation error estimate, Eq.~(\ref{eq:TruncError}) for the same data.
When the junk radiation is at its peak, around $t=150M$, the truncation is as high as $-2.5$ (i.e. the actual representation
error can be as large as $10^{-2.5}$).  After the junk radiation passes through, the truncation error settles to a much more
desirable value, around $-6$. This is consistent with the six orders of magnitude covered by the unfiltered coefficients
at $t=500M$, as shown in the upper panel.}
\end{figure}

Pile-up modes are identified by a local convergence factor estimate, involving the mode under consideration and its next
few higher neighbors.   Once again, the AMR algorithm benefits from a pile-up mode counter that is
a continuous function of the data (rather than a discrete counter).  In order to measure the extent to which a given mode
$j$ is a to be considered as a pile-up mode, we measure a local convergence factor around the $j$-th point,
\begin{equation}
\tilde {\cal C}_j \equiv - {\cal S}[j,\min(\tilde N_k,j+4)] ,
\end{equation}
and compare it to the overall convergence factor, ${\cal C}[P_k]$.
The number of pile-up modes is then defined to be
\begin{equation}
{\cal P}[P_k] \equiv \sum_{j=2}^{\tilde N_k -1 } 
         \exp
           \left[
                -32 \left( 
                    \frac{ \tilde {\cal C}_j}{{\cal C}[P_k] }
          \right)^2 
          \right] .
\label{eq:PileUpMode}
\end{equation}
If the local convergence factor estimate $ \tilde {\cal C}_j $ associated with the mode $j$ has a value close to that of the
overall convergence factor, i.e. if
$$
  \tilde {\cal C}_j \approx {\cal C}[P_k] ,
$$
than that mode will have an $O($roundoff$)$  contribution to ${\cal P}[P_k]$.  If it is near zero, i.e., 
$$
   \tilde {\cal C}_j   \ll {\cal C}[P_k]
$$
this indicates that we are down in the noise floor
of the power-monitor plot, and the contribution of the individual mode to the number of pile-up modes will be $O(1)$.   
A plot showing the number of pile-up modes, and the associated power monitor, for a typical BBH production run is shown in
Fig.~\ref{fig:TruncErrAndPileUp}.

A third quantity essential in our spectral AMR algorithm is the truncation error estimate associated with the power-monitor $P_k$.
This is computed using the expression
\begin{eqnarray}
{\cal T}[P_k] &\equiv& \log_{10}\left(\max\left(P_1,P_2\right) \right)
- \frac{ \sum_{j=1}^{\tilde N_k} \log_{10}\left(P_j\right) w_j } {  \sum_{j=1}^{\tilde N_k} w_j }, \quad \mbox{where}
\\
\quad w_j &\equiv& \exp\left(-\left(j-\tilde N_k+\frac{1}{2}\right)^2\right) .
\label{eq:TruncError}
\end{eqnarray}
The aim here is to count the number of digits resolved by the given set of spectral coefficients.
This is computed as the difference 
between the power in the larger of the two lowest order modes, $\log_{10}\left[\max\left(P_1,P_2\right) \right]$, and the
power in the highest modes, which itself is computed as an exponentially weighted average, giving maximum weight to the
last two points (refer to Fig.~\ref{fig:TruncErrAndPileUp}).

The truncation error ${\cal T}[P_k]$, the convergence factor  ${\cal C}[P_k]$ and the number of pile-up modes ${\cal P}[P_k]$
are the three essential measures that our AMR algorithm relies upon.

\section{Adaptive Mesh Refinement}

In finite-difference based discretization schemes one commonly used AMR algorithm is that described by
 Berger and Oliger in Ref.~\refcite{Berger1984}. In such a scheme the evolution would be done  on a sequence
of refinement levels simultaneously.   The truncation error between the two highest refinement levels can then
be used to monitor the accuracy of the evolution and to dynamically assign the number of refinement levels 
to a particular region of the grid.  Each refinement level
will have a factor of two smaller grid-step (and possibly time-step) than the next coarser level. 
For a region that has
$N$ refinement levels at a particular instant of time, there will simultaneously co-exist the highest level, the coarser one, the next
coarser one, to the coarsest, all used to discretize this region at varying levels of accuracy.

Our approach to mesh refinement is different.  The {\tt SpEC} code uses  a single layer of grid for a  given (BBH) simulation.
On a very basic level, our mesh refinement algorithm monitors the truncation error estimates associated with each irreducible
topology, within each subdomain, throughout the run and then
adjusts the number of collocation points (and, accordingly, the number
of spectral basis elements) in order to keep the local truncation error within some desired range.  This is called $p$-type mesh
refinement.

In addition, as the objects approach each other while on their trajectories, our control system and the associated
maps shift the boundaries of the various subdomains with respect to each other in order
to accommodate the requirement that the
excision boundary must stay inside the apparent horizons at all times.   As a consequence, from time to time subdomain
 boundaries need to be
re-drawn  in order to reduce the stretching or compression of the grid
caused by their continuous drifting under the action of the various maps.
When this change of the grid structure implies splitting or joining of subdomains, we call this as $h$-type mesh 
refinement.
When it preserves the number of subdomains but shifts their boundaries, it is  called $r$-type mesh refinement.
In addition, our AMR driver splits subdomains when the number of spectral modes required for a given target accuracy
exceeds some threshold.  This happens regularly with the spherical shells surrounding the smaller black hole in 
high mass ratio mergers, and is another example of $h$-type mesh refinement in our code.
   In the following sections we give a description of each of these elements of our AMR algorithm.

\subsection{$p$-type mesh refinement}

\begin{figure}
\centerline{\includegraphics[scale=0.4]{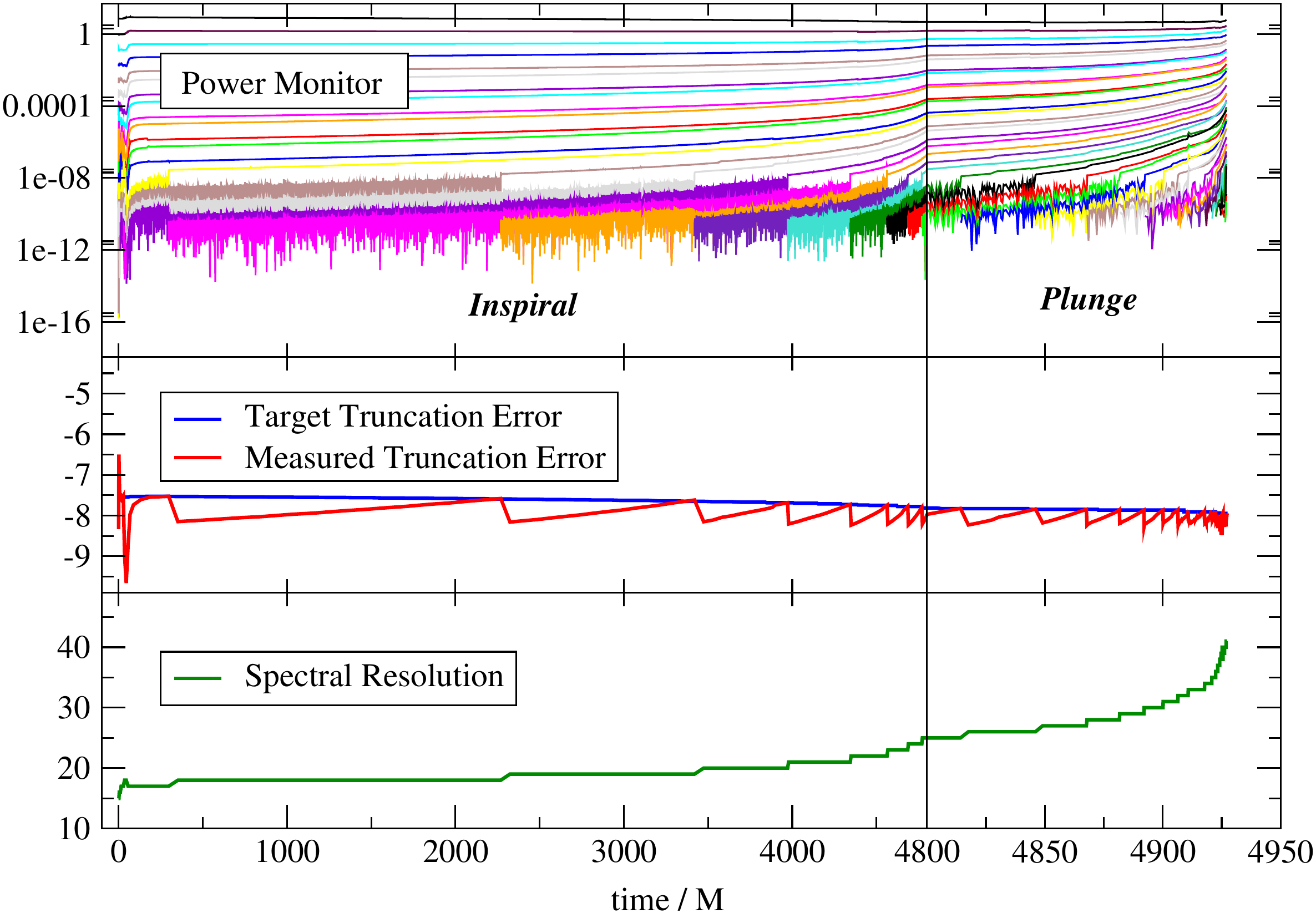}}
\caption{\label{fig:AmrB0Shell} (Color online) The plot illustrates $p$-type mesh refinement for the innermost spherical shell around
the smaller black hole (labeled {\tt SphereB0}), for a $q=7.1875$ non-spinning BBH inspiral and merger. 
 The { top} panel shows the combined scalar
spherical harmonic power-monitor Eq.(\ref{eq:YlmPowerMonitor})  for
the variables $\Phi_{iab}$ and $\Pi_{ab}$ as a function of time.  The lowest four noisy modes 
are filtered (set to zero) before each time-step.  The power monitor is evaluated after the time-step, showing that
the coefficients stay small from one step to the next. 
 The { middle} panel shows the target truncation error (blue)
as well as the actual numerical truncation error (red). The { lower} panel shows the spectral resolution
(in this case $L_{\max}+1$) as a function of time. Each time the measured numerical truncation error
goes above the target value, the AMR driver adds a spectral basis element to the representation (i.e., increments
 $L_{\max}$ in the $Y_{lm}$ expansion), bringing the truncation error to below its target.}
\end{figure}

To zeroth order the $p$-type mesh refinement algorithm monitors the truncation error estimate ${\cal T}[P_k]$ for each 
power-monitor $P_k$ associated with the individual subdomains, and adjusts the number of collocation modes (by adding
or removing modes) such that the accuracy of the spectral representation of the evolved quantities is within some desired range.
This behavior is seen on Fig.~\ref{fig:AmrB0Shell}.  The algorithm behind the plot contains additional elements described
in the remainder of this section. Nevertheless, the plot is a good illustration of how $p$-type mesh refinement works in our code.

\begin{figure}
\centerline{\includegraphics[scale=0.4]{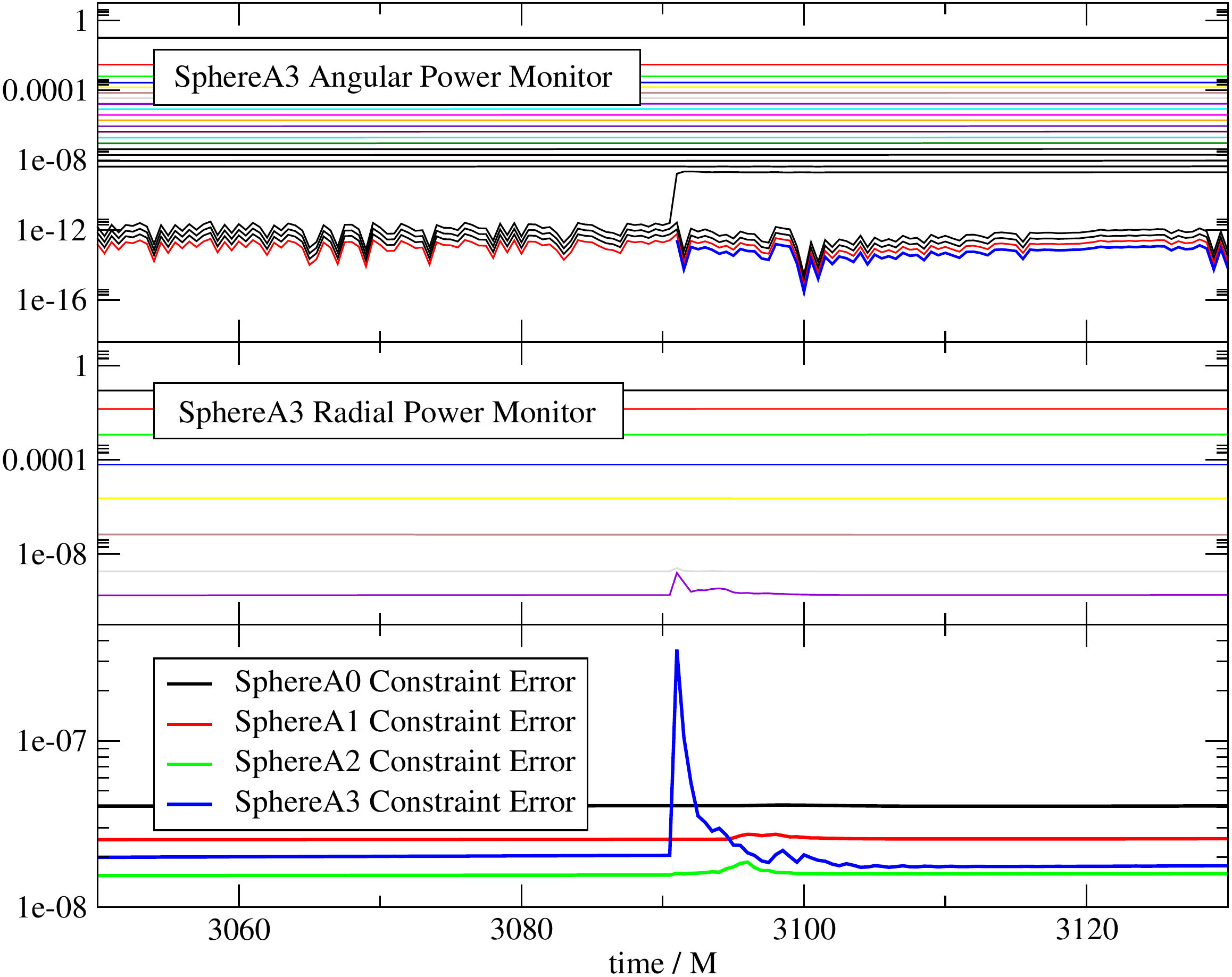}}
\caption{\label{fig:AmrA3Noise} (Color online)
Spherical and radial power-monitors as well as constraint errors during a portion of
 a $q=7.1875$ non-spinning BBH simulation, displaying the effects of a change in resolution.
{ Top panel}: Angular power-monitor for the outermost  spherical shell (labeled `{\tt SphereA3}`)
around the larger black hole.   At around $t=3090$ the AMR driver increases the angular resolution
$L_{\max}$ from $22$ to $23$.  One implication is that the $L=18$ coefficients, 
which had been the lowest of the modes filtered 
according to
 Eq.~(\ref{eq:HeavisideFilter}), 
will now be the highest of the un-filtered modes.
The power-monitor shows this coefficient joining
the evolved (or un-filtered) modes, 
while the newly added $L=23$ mode will be the highest of the four filtered modes.
{ Middle panel}: Radial power-monitor of {\tt SphereA3}.  At the time the angular resolution
is increased, the highest coefficient (which is the most sensitive as being the smallest in value)
shows a temporary spike of $O(10^{-9})$.  This spike is not dictated by the evolution equations,
but rather is a consequence of the sudden change in the numerical algorithm (i.e., a change in the grid).
{ Lower panel}: The $L_2$ norm of the constraint error in the spherical shells around
the larger black hole.  At the time of the angular resolution change of {\tt SphereA3}
there is a small, $O(10^{-7})$ spike seen in that same subdomain. A much smaller, 
$O(10^{-8})$ spike is seen in the neighboring ${\tt SphereA2}$ a short time later.
This constraint error injection can be seen as a consequence of the sudden change
of the evolution equation prescribing the $L=18$  coefficient resulting in a jump
from near roundoff values to $O(10^{-9})$.
This sudden change, dictated by the numerical scheme rather than Einstein's equations,
generates a constraint violating mode that dissipates away on a time-scale of $O(10)M$.
}
\end{figure}

Each time a basis associated with a particular subdomain is extended, we find that
the numerical evolution system needs
some time to adjust and find a new quasi-equilibrium for the evolved coefficients.  This can be seen both in the power-monitors
and in the constraints, as shown on Fig.~\ref{fig:AmrA3Noise}. 
For this reason, our current algorithm calls the AMR driver at pre-set time-intervals (rather than every time-step).  Each time
a particular grid extent is changed, it will be kept at the new level for some time before it is allowed to change.  During the early inspiral,
grid extents are changed  no more frequently than every $O(100)M$ evolution time. This time-interval is gradually decreased and
during the plunge individual grid extents are allowed to change every $O(1)M$.    An alternative way of dealing with this
same problem would be to use a smooth filter for all spectral expansions and control the parameter of this filter as a continuous function of time so that when the AMR driver decides to remove a point the filtering coefficients gradually become
stronger for the highest unfiltered coefficient, 
until it reaches $O($roundoff$)$ level. 
At this point
the highest coefficient could be safely be removed (as it could only gain power from coupling to the
next highest coefficient which by now is also filtered) 
and doing so would not present a noticeable
 `shock' to the numerical evolution scheme.
A similar approach could be applied for adding a new spectral coefficient.   
Implementation of such a  `smooth' resolution change
is future work.

\subsubsection{Target truncation error}
\label{sec:TargetTruncError}

An important element in the robustness of our current AMR algorithm is the proper setting
of the target truncation error.
An essential requirement on the angular resolution of the spherical shells  is that they must be able to represent
the data at a level accurate enough for the needs of the horizon finder.   The horizon finder, however, does not depend
only on the angular representation of the data.  Radial resolution is required as well.  Another important physical aspect  that we want
to represent accurately in our simulation is the amplitude and, more importantly, the phase of the  gravitational wave (as gravitational
wave detectors are more sensitive to phase than amplitude).  In the outer layers of the grid, these waves propagate as
outgoing spherical wavefronts, which do not require a lot of resolution.  It is important, however, to get the orbital phase of
the binary right, since if the orbits are inaccurate then the waves (generated by the motion of the black holes) cannot be accurate either.
For the sake of simplicity, we use the label `A' for the black hole contained in the subdomains {\tt SphereA}$n$ 
and `B' for the black hole contained in the subdomains {\tt SphereB}$n$.
In the region immediately surrounding the black holes (including the subdomains discretizing their gravitational
attraction between the two excision boundaries),
we find that we need higher accuracy. Our target truncation
error function is written as 
\begin{equation}
\label{eq:TruncErrorFunction}
{\cal T}^{\max}\left[ w_A,  w_B\right] \equiv {\cal T}_0 - 4
\frac{ w_A +   w_B }{w_A+w_B+1}  \; , 
\end{equation}
where the weighting coefficients $ w_A,w_B$ are the $L_\infty$ norm on each subdomain
of the smooth weight functions  
$$\tilde w_X(x^i,t), \quad X=A,B$$
which are expressed in terms of inertial coordinates $x^i$,
with maximum values around the time-dependent coordinate-center $x^i_X(t)$ of
the individual black holes labeled `A' and `B':
\begin{eqnarray}
&& \tilde w_X(x^i,t) \equiv \exp\left[ - \sum_i \left(\frac{x^i-x_X^i(t)}{c_X}\right)^2 \right], \quad X=A,B, \quad \mbox{and}
\\
&& 
c_A \equiv \sum_i  \left. \left( x_B^i(t)\right)^2\right|_{t=0}
, \quad
c_B \equiv \sum_i  \left. \left( x_A^i(t)\right)^2\right|_{t=0} .
\end{eqnarray}

\begin{figure}
\centerline{\includegraphics[scale=0.5]{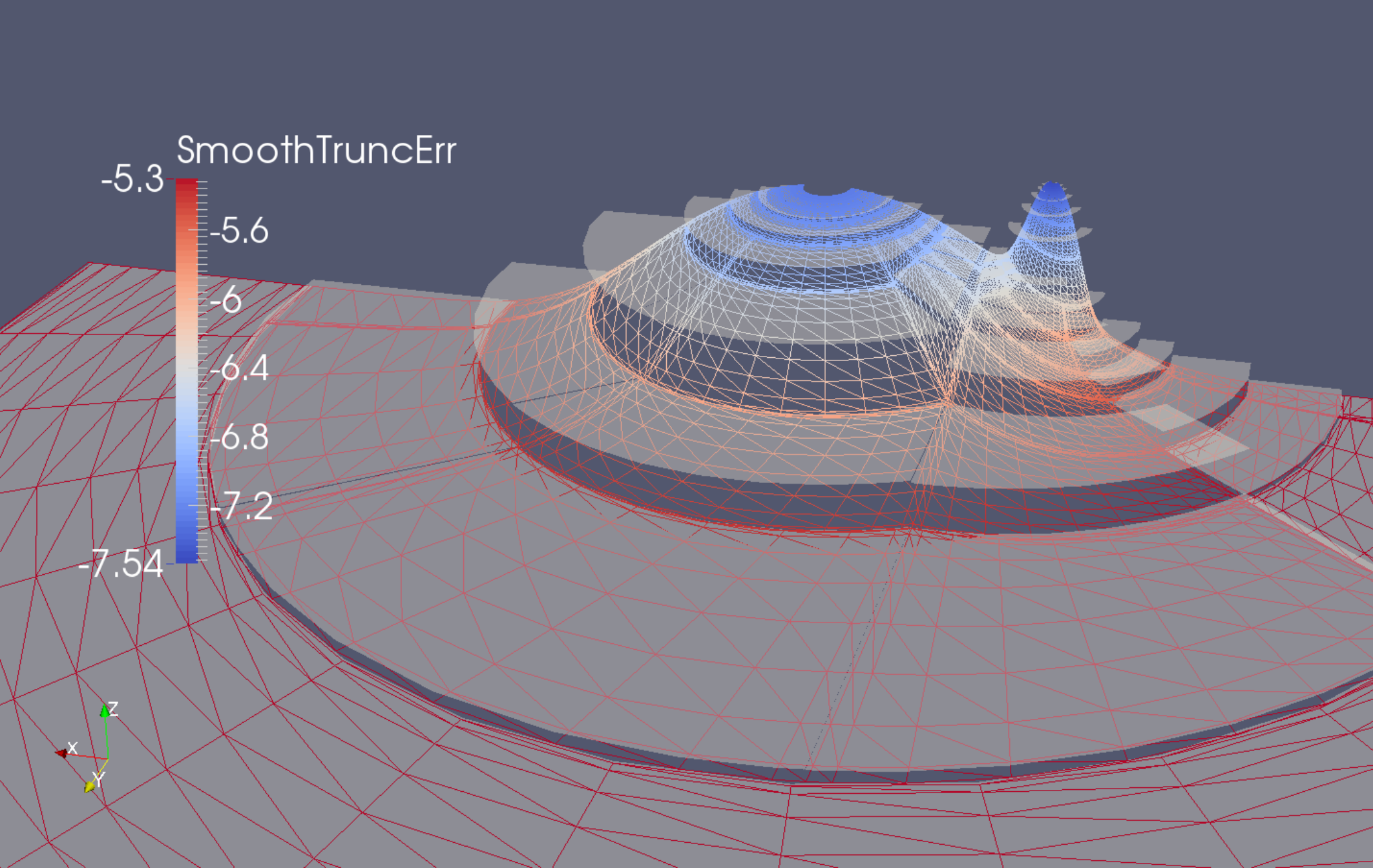}}
\caption{\label{fig:TruncErrorOnDomain} (Color online) A $z=0$ half-slice of the target truncation error
${\cal T}^{\max}\left[ w_A,  w_B\right]$ (solid gray) , shown on the subdomains near the black holes, as well as
its smooth approximate ${\cal T}^{\max}\left[ \tilde w_A(x^i,t), \tilde w_B(x^i,t)  \right]$, evaluated at $t=0$
 (color-coded wireframe).
The plot shows
 the start of a non-spinning, mass ratio $7.1875$ BBH simulation. 
The individual black holes are initially located on the $x$ axis. The elevation
of the surfaces is set by the magnitude of ${\cal T}$, so that higher elevation means more accuracy.
Near the individual black holes the truncation error requirements are stricter, as shown. }
\end{figure}

Note that here we use the initial distance of black hole `A' to the coordinate origin to set the falloff rate of the
Gaussian controlling the  truncation error requirement around black hole `B', and vice-versa.  
The rationale is that if one black hole is
larger, it starts off closer to the origin, i.e., the other smaller black hole will have a smaller Gaussian around it.
As a further constraint on
 ${\cal T}^{\max}\left[ w_A,  w_B\right]$, this quantity cannot differ by more than $\log_{10}2$ between two neighboring
subdomains. This rule is enforced by setting stricter truncation error on those subdomains that otherwise
would end up being coarser than intended.  We plot both the subdomain-wise constant  ${\cal T}^{\max}\left[ w_A,  w_B\right]$
and the smooth function $ {\cal T}^{\max}\left[ \tilde w_A(x^i,t), \tilde w_B(x^i,t)  \right]$ in Fig.~\ref{fig:TruncErrorOnDomain}.

\subsubsection{Special rules for spherical shells}
\label{seq:AmrRulesForShells}

In order to control the accuracy of the apparent horizons themselves (and, as such, of the numerically evolved black holes),
we define three criteria that factor into the final choice of the angular resolution of the horizon finder:
\begin{enumerate}
\item 
We form a $Y_{lm}$ expansion of the shape function of the horizon surface, $r(\theta,\phi)$ which is defined as the
coordinate distance between the point $(\theta,\phi)$ on the surface and its coordinate center.%
\footnote{Note that the angular 
coordinates are defined local to the horizon center and are independent of where the horizon is located with respect
to the global coordinate-systems center.}  Given this spectral expansion, we form a  
power monitor Eq.~(\ref{eq:YlmPowerMonitor}) and then compute the  truncation error estimate 
Eq.~(\ref{eq:TruncError}) and the number of pile-up modes Eq.~(\ref{eq:PileUpMode}).  If the surface shape is
coarser than required and if the number of pile-up modes is not larger than some threshold, we
increase the accuracy of the horizon finder.
If the number of pile-up modes is already at its limit, adding more modes would not help.
\item
As a second method of determining adequacy of resolution, we form a $Y_{lm}$ expansion of the
 areal radius of the apparent horizon surface and require that it be represented with a truncation error that is within
a desired range.   Note that the surface areal radius will reduce to the coordinate radius in flat space time.  Imposing
an accuracy requirement on the surface areal radius requires an accurate representation of the metric quantities
on the horizon surface.  This in turn allows for accurate measurement of physical quantities such as areal mass and spin.
Similar to the coordinate radius, the horizon finder will increase accuracy if the areal radius is under resolved 
(and does not have too many pile-up modes).
\item
As yet another method for imposing an accuracy requirement on the horizon finder algorithm, we monitor the
residual of the horizon finder.   If larger than requested, the angular resolution is increased.
\end{enumerate}
If the accuracy of the surface is better than required by any of these measures, the horizon finder will decrease accuracy.

The angular resolution of the horizon finder can be used to set a lower limit on the angular resolution of the underlying
three-dimensional evolution grid.   Given the top-4 Heaviside filters Eq.~(\ref{eq:HeavisideFilter}), we request that the set of 
spherical shells around the excision boundary associated with a given horizon 
must have five additional $L$ modes beyonf the surface finder's resolution. This
implies that, should any of the horizon finder accuracy criteria request higher resolution, additional data is
available upon interpolation from the evolution grid.  Once the horizon finder increases its resolution, the spherical
shells around its excision boundary have their angular resolution increased.

The angular resolution of the horizon finder is a useful lower limit for the spherical shells around an excision
boundary.  There are additional, heuristic rules that we found useful not only for accuracy but also for 
well behaved constraints.

The first such rule is that given the angular resolution $L$ of a spherical shell subdomain, its neighbors must not have
angular resolutions larger than $L+1$ or smaller than $L-1$.     There is yet another heuristic rule relating the
angular resolution of the excision subdomain to the spherical shells surrounding it.
As the binary system evolves, the further a subdomain is from the excision boundary, the less spherical symmetry
will be in the metric data. Based on truncation error estimates, this means that the AMR driver will
assign larger angular resolution to the spherical shells that are further from their associated excision 
boundary and closer to the other black hole. In certain cases, however, the metric data on the excision boundary may
itself require a large resolution (e.g., if the black hole has a very large spin or distortion). When
the subdomain next to the excision boundary has a larger angular resolution that its neighbors,
we find that  the constraint errors
grow on a short time-scale. This may be related to the fact that if a given subdomain boundary has subdomains of
different resolution on its two sides,  those angular modes that are represented on only one side of the boundary will
be reflected. When the excision subdomain needs high resolution, it may also be responsible for a larger amount
 of high frequency noise generation.  If this is the case and its neighbor has a lower angular resolution, 
all that noise can get trapped in the excision subdomain and
have non-desired effects on the black hole itself.  
Thus, to rule this out, if the excision subdomain needs resolution then
all its neighbors are given the same, or larger, resolution.

The angular resolution of the set of shells resolving the wave-zone is driven by a much simpler set of rules.  In this region 
we base the angular resolution of the shells on the truncation error estimate Eq.~(\ref{eq:TruncError}) derived from
the  power-monitor associated with the scalar $Y_{l,m}$ expansion of the main evolution variables 
$\left\{\psi_{ab},\Phi_{iab},\Pi_{ab}\right\}$.  Similar to the inner sets of shells,we do not allow neighboring wave-zone
shells to have their angular resolution differ by more than one $L$.  As a possible future improvement,
it may be worth forming a power-monitor from the spherical harmonic
decomposition of the Weyl scalar $\Psi_4$ and require that the truncation error associated with this power-monitor
also be within limits.

\subsubsection{Pile-up mode treatment}

An essential element of our Spectral AMR algorithm is the treatment
of pile-up modes.  The source of these modes is not fully understood,
although we expect that inadequate filtering or
unresolved high frequency modes injected
through a subdomain boundary can lead
to such modes.   It is also expected that, when the evolved quantities
have a lot of non-trivial features, a more fine-grained definition
of the truncation error estimate Eq.~(\ref{eq:TruncError}),
convergence factor Eq.~(\ref{eq:ConvFactor})
and pile-up modes Eq.~(\ref{eq:PileUpMode}) could
prove helpful in better tracking and controlling the
numerical representation of the evolved quantitites.

As seen in Fig.~\ref{fig:PowerMonitor}, the presence of these pile-up modes does not contribute to the accuracy of the data.
On the contrary, we find in practice that these modes are constraint violating modes.
This suggests that pile-up modes are result of the numerical approximation
and not a property of the underlying analytic system.  In other terms,
Einstein's equations will likely not dictate the presence of such modes.
Rather, these develop on the grid as a result of limitations of our
particular numerical scheme.  We do not have experience with other
spectral evolution codes and therefore are unable
to assess whether these modes are a generic property of spectrally
evolved partial differential equations. In our experience, however,
the less pile-up modes, the better.
For this reason we make it a priority to eliminate them, or to prevent their creation.

\begin{figure}
\centerline{\includegraphics[scale=0.4]{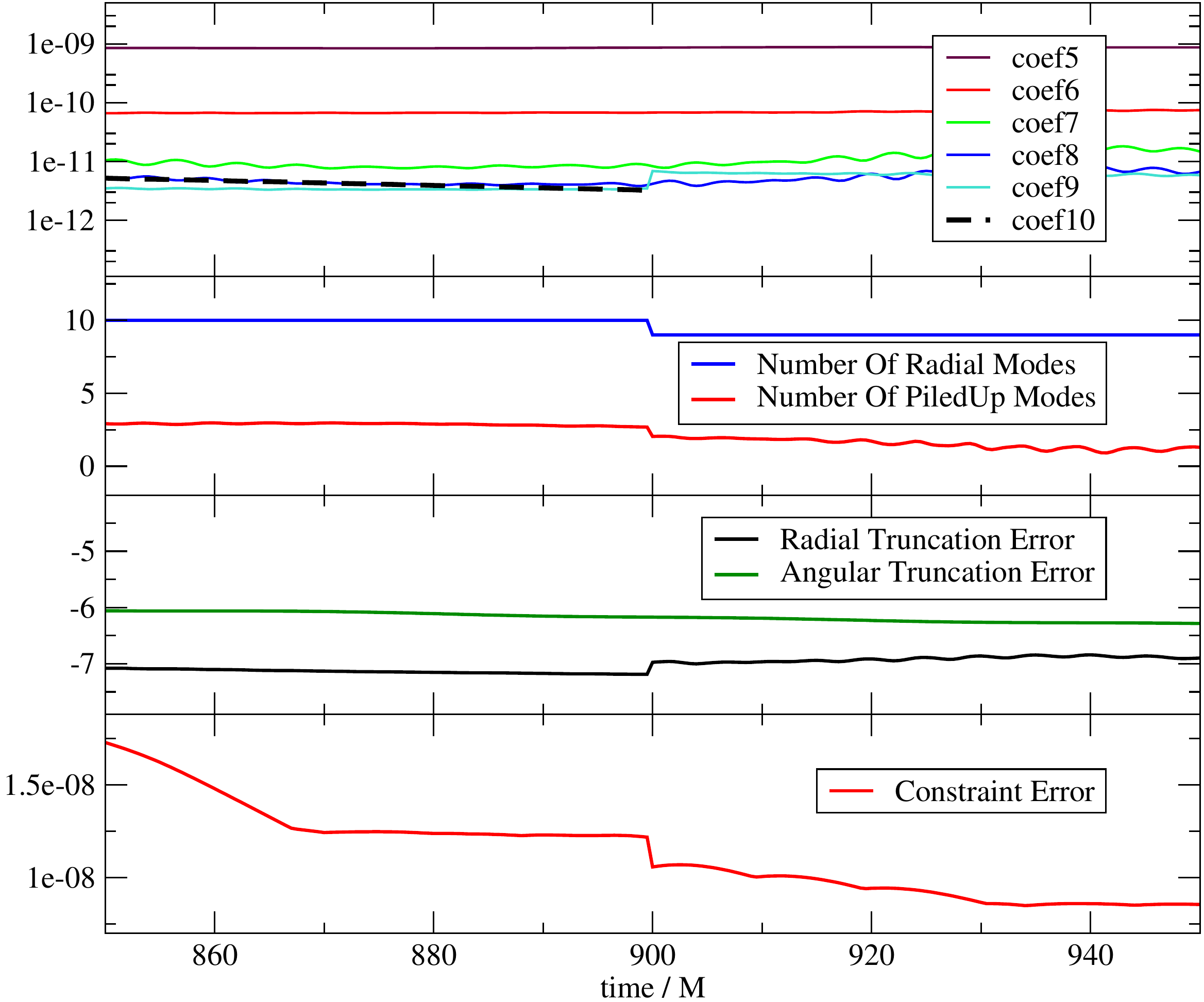}}
\caption{\label{fig:SphereC6_PileUpMode} (Color online)
Pile-up mode removal and its effects in a non-spinning $q=7.1875$ BBH evolution.
The { first} panel from the top shows the highest order modes in the
 radial power-monitor of the wave-zone spherical shell labeled {\tt SphereC6},
 for the time-interval $[850,950]$.
At around $t=900$ the AMR algorithm removes a radial point in order to reduce
the number of pile-up modes in this subdomain.  The { second} panel
from the top  shows that
the pile-up mode diagnostic Eq.~(\ref{eq:PileUpMode}) is indeed decreased as
the number of radial modes decreases by one. The { third} panel from the top displays
the truncation error Eq.~(\ref{eq:TruncError}) associated with both the
angular representation (green curve) and the radial representation (black curve).
As seen from this panel, the overall truncation error in this subdomain is dominated by
the angular representation -- removing a radial mode should have little effect on
overall accuracy. The { bottom} panel plots the $L_2$ norm of the (unnormalized)
 constraint error in the same subdomain.  
Remarkably, the constraints
drop at around $t=900$ by a sizable factor, as a pile-up mode is removed. This
confirms that the pile-up mode being removed was a constraint violating mode.
We find, in general, that removing these modes
 can in fact lead to a lower amount of constraint violation on the grid.
}
\end{figure}

One immediate way of handling pile-up modes is to remove them. If a power-monitor shows these modes, we simply
reduce the grid-extent associated with the piled-up power-monitor.   In some cases doing so would violate some of
our other
rules of thumb (e.g., the horizon finder may need a certain minimum resolution).   In this case we keep
the resolution of neighboring subdomains from increasing until the pile-up modes leave the grid either by filtering or by propagating away from the subdomain under consideration.  Fig.~\ref{fig:SphereC6_PileUpMode} illustrates, what we find
in most cases, that removal of pile-up modes helps reduce the constraint violating modes on the grid.
This suggests that
these modes do not result from the underlying analytic system but are instead a numerical artifact.

\subsubsection{AMR driven by projected constraint error}

As an alternative way of determining whether a given grid extent needs to be changed, one can monitor
the constraint quantities ${\cal C}_{ia}, {\cal C}_{iab}$, defined in Eqs.~(\ref{eq:Cia}),(\ref{eq:Ciab}).
The first of these indexes are related to derivatives, while the rest of the indexes are related to the 
evolution frame  of the main system (i.e., the inertial frame).   In order to derive an accuracy
requirement based on the constraints, we make use of the fact that each subdomain is
constructed as a product of topologies (see Eq.~(\ref{eq:TopProduct}) for a list of these topology products).
We form an estimate of the contribution of the truncation error associated with the individual topologies
by projecting  the derivative indexes from the evolution frame
$\{x^k\}$ into the frame associated with
the topology product that is used to construct the grid $\{\hat x^{\hat k}\}$. 

That is, for a spherical shell, we project
the Cartesian $i$ index of  ${\cal C}_{ia}$ onto spherical coordinates, summing over the non-derivative indexes.
In general, we write
\begin{eqnarray}
    {\cal E}^P_{\hat k} \left[ {\cal C}_{ia} \right] &:=& \frac{1}{N_{\hat k}}
\sqrt{\sum_a \left( \sum_i \frac{\partial x^i}{\partial \hat x^{\hat k}} {\cal C}_{ia}\right)^2  }
\\ 
    {\cal E}^P_{\hat k} \left[ {\cal C}_{iab} \right] &:=& \frac{1}{N_{\hat k}}
\sqrt{\sum_{ab} \left( \sum_i \frac{\partial x^i}{\partial \hat x^{\hat k}} {\cal C}_{iab}\right)^2  }
\end{eqnarray}
where the normalization coefficient $N_k$ is set using
\begin{equation}
N_{\hat k} = 
\sqrt{\sum_i \left( \frac{\partial x^i}{\partial \hat x^{\hat k}} \right)^2  }.
\end{equation}
The $L_2$ or the $L_\infty$ norm  of 
${\cal E}^P_{\hat k}[{\cal C}_{ia}]$ and  
${\cal E}^P_{\hat k}[{\cal C}_{iab}]$ over the individual subdomains
can then serve as an indicator of whether the spectral expansion associated 
with the topological coordinate $\hat x^{\hat k}$ has an adequate accuracy.  We have tested this algorithm in conjunction
with the truncation-error based AMR described in Sec.~\ref{sec:TargetTruncError}, always applying the stricter
of the two requirements.  We find that this additional accuracy requirement can be helfpul in the more dynamic parts
of the BBH simulation (such as during plunge) where constraint violating modes can develop faster and spoil the 
physical properties of the system.

\subsection{$h$-type mesh refinement}

In a general spectral AMR algorithm, $h$-type mesh refinement is motivated largely by efficiency considerations.
If it takes too many spectral coefficients to resolve data along a particular axis of the subdomain then differentiation may
become very expensive, the time-step may be limited due to a stricter CFL limit, memory limitations may arise,
and load balancing may become a problem if the application is parallelized.  All of these concerns are
relevant for our code as well.  
We monitor the convergence factor Eq.~(\ref{eq:ConvFactor})
of a given power-monitor and when it reaches values below some arbitrary threshold, the subdomain will be split.
In our binary black hole evolutions typical values for the convergence factor are of order unity, so our threshold
for splitting is set to $0.01$.   In addition, we find it useful to set hard limits on the maximum number of spectral coefficients
allowed for a specific irreducible topology of a give subdomain type.  For instance, our rule
is not to have more than 20 radial points in any spherical shell.  If the $p$-type AMR driver
finds that this number is not sufficient then the subdomain will be split in the radial direction.\footnote{One such 
scenario, found in a number of our binary black hole inspirals, is the radial splitting of some of the outermost spherical
shells.  This results from the action of the cubic scale map $\map{Scaling}$, which is responsible for
shrinking the interior of the grid as the binary approaches merger, while some of the outer subdomains get stretched.
The AMR driver monitors this through the truncation error of the radial spectral expansion and keeps adding
coefficients until the user-specified limit of the maximum allowed radial collocation points is reached. 
Then the subdomain gets split and the evolution can proceed.}

Another role of the $h$-type AMR in our runs is around the individual excision boundaries.  For a number
of our binary simulations one of the apparent horizons will pick up more and more angular resolution, forcing the
underlying spherical shells to add spherical harmonic coefficients with higher and higher values of $L$.  When
the angular resolution reaches a user-specified limit, the spherical shell is split.  We find it important that when
such a spherical shell is angularly split, all shells outside it (among those around the same excision boundary) 
be split as well.  This, once again, is a rule we found useful by trial and error
when optimizing for small constraint violation.

\subsubsection{Shell-dropping around the excision boundaries}
\label{sec:ShellDrop}

What we call our  `shell dropping algorithm'  is a use of $h$-type AMR which is an absolute must in our binary evolutions.

\begin{figure}
\begin{center}
\includegraphics[scale=0.48]{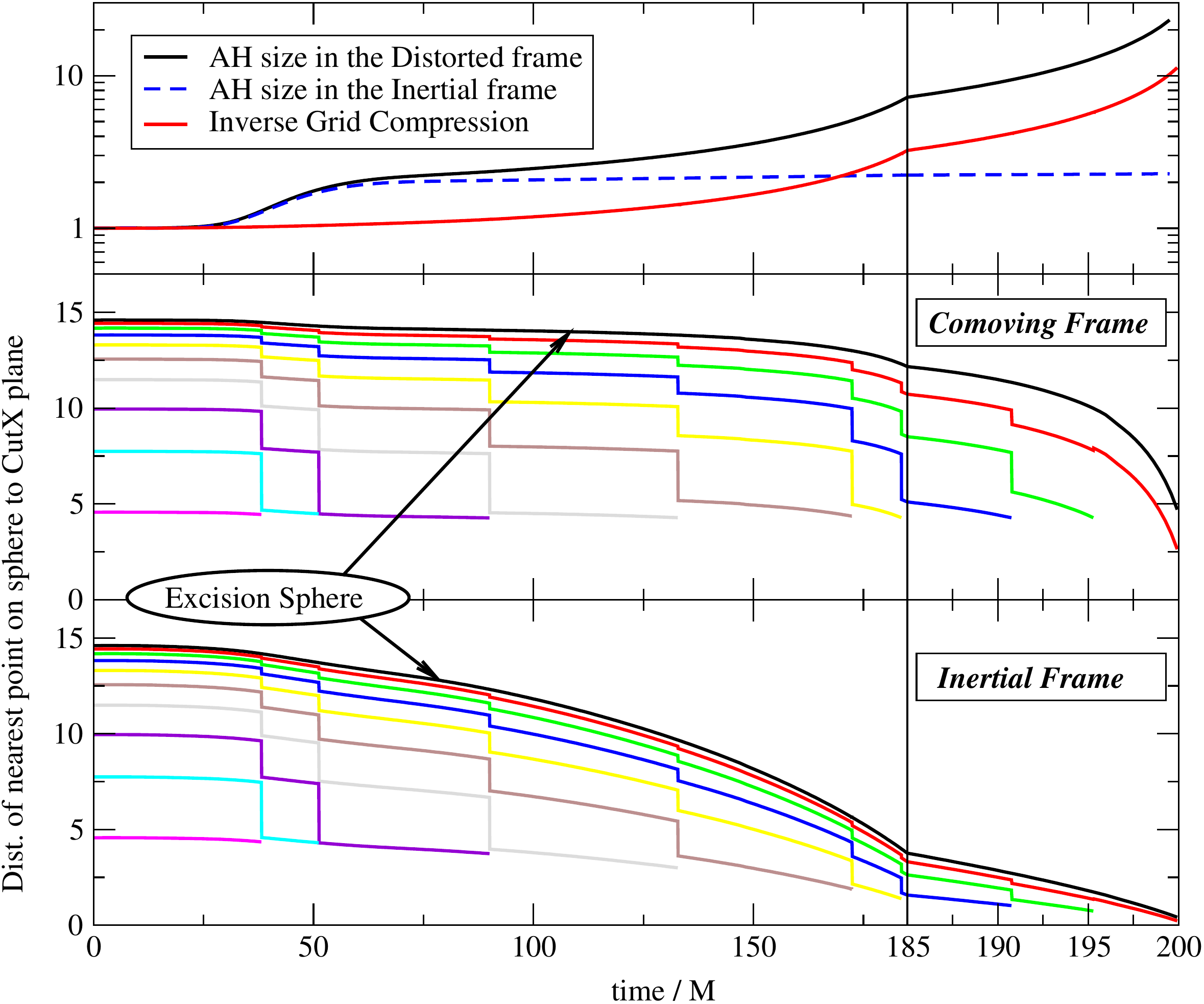}
\end{center}
\caption{\label{fig:ShellDrop}
(Color online) This plot illustrates our shell dropping algorithm
for an equal mass non-spinning head-on simulation, where the individual black holes
are initially centered at $x=\pm 15,y=z=0$ in coordinate units. 
The {upper panel} shows the apparent horizon size in inertial (blue) and
co-moving (black)  coordinates as well as the inverse of the grid compression factor
(called `expansion factor' in our code).  As the black holes approach each other,
the distance between their centers (in inertial coordinates) decreases.  In order
to keep the excision boundaries concentric with the black holes, the inner portion
of the grid is compressed.  In this compressed coordinate system the black hole
looks larger (a coordinate effect).  The excision boundary is expanded, accordingly,
in order to keep it near the black hole's horizon.  
The { middle panel}
shows the co-moving frame coordinate distance
between the various spherical boundaries around one of the black holes
and the {\tt CutX}-plane, placed at $x=0$ for this simulation.   
As the excision boundary radius grows in the co-moving frame,
the spherical shells around it
grow as well, with the distance between the outermost such shell and the
{\tt CutX} grid-plane decreasing.  Periodically, when this distance falls below a certain threshold,
the number of spherical shells is reduced by one and the radii of the remaning shells get adjusted.  
At around time $\approx 196M$, when
only two spherical shells remain, a thin spherical shell is created just thick enough 
to contain the interpolation stencils for the horizon finder, 
but otherwise small in order to delay the outer boundary of this shell coming
too close to the {\tt CutX}  grid-plane.  The { lower panel} shows the inertial-frame
coordinate distance between the various spherical shells around the excision boundaries and the
{\tt CutX} plane.
  The radius of the excision sphere in this frame is related
to the inertial coordinate radius of the black hole and as such it stays roughly constant
during the simulation.  However, in this frame there is no mapping to keep the center
of the black holes at a fixed position, so that the excision sphere (and the surrounding
shells) approach the grid-plane as the BBH system nears merger.   This, in effect, leads to a higher resolution
grid during plunge, which helps in preserving accuracy.  Please refer to Fig.~\ref{fig:ShellDropVtk}
for more details.
}
\end{figure}

\begin{figure}
\centerline{
\includegraphics[scale=0.19]{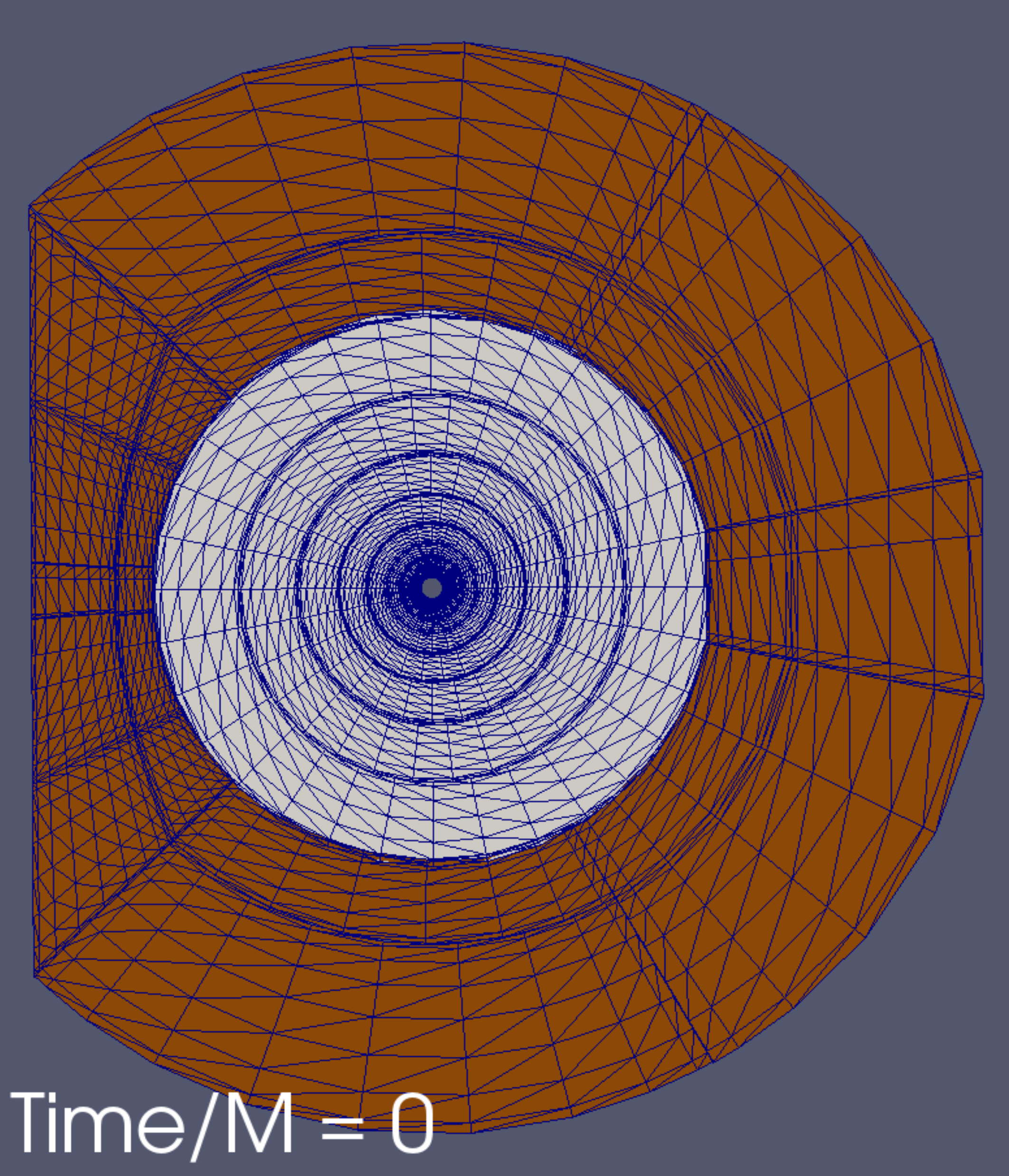}
\includegraphics[scale=0.19]{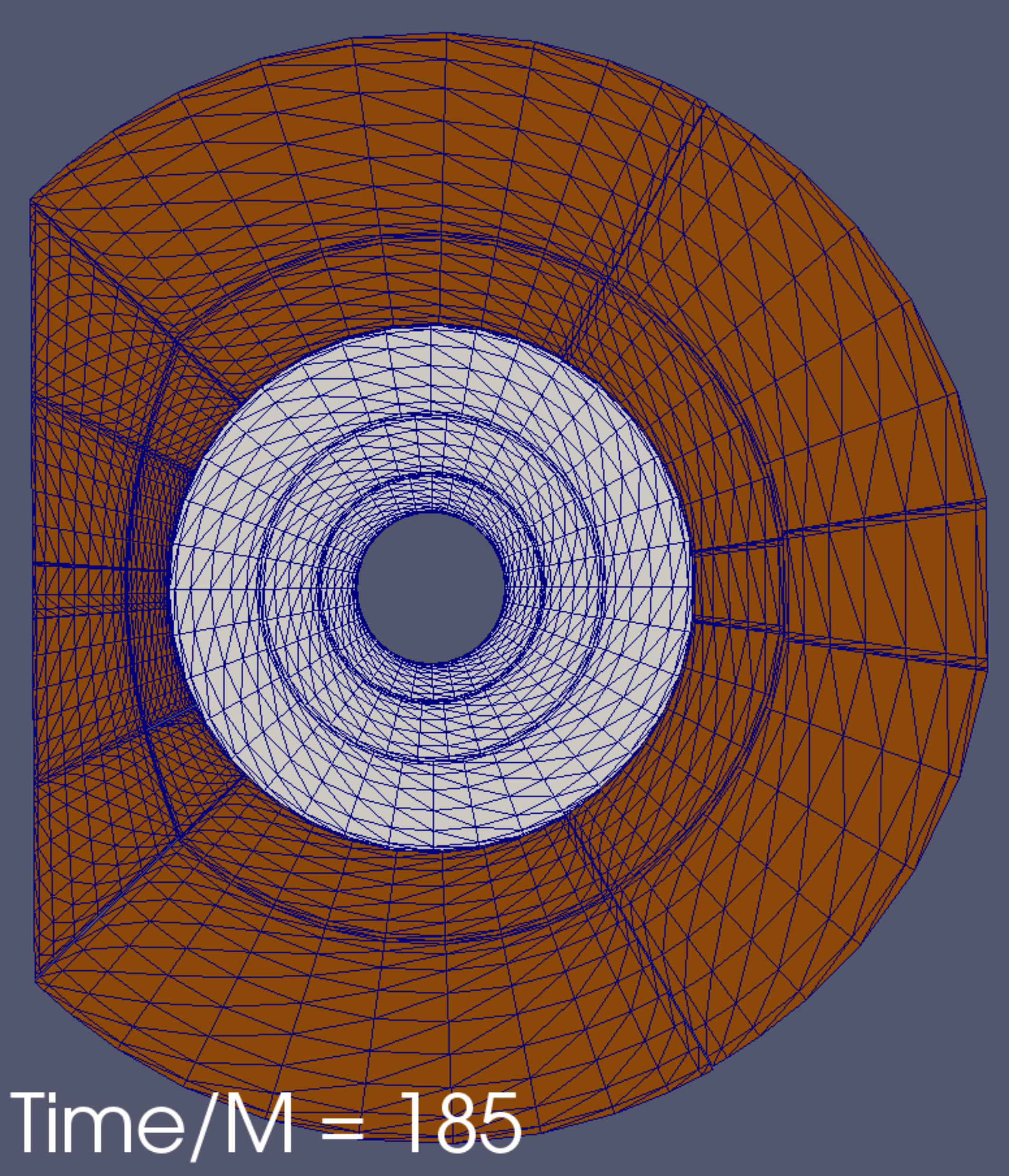}
\includegraphics[scale=0.19]{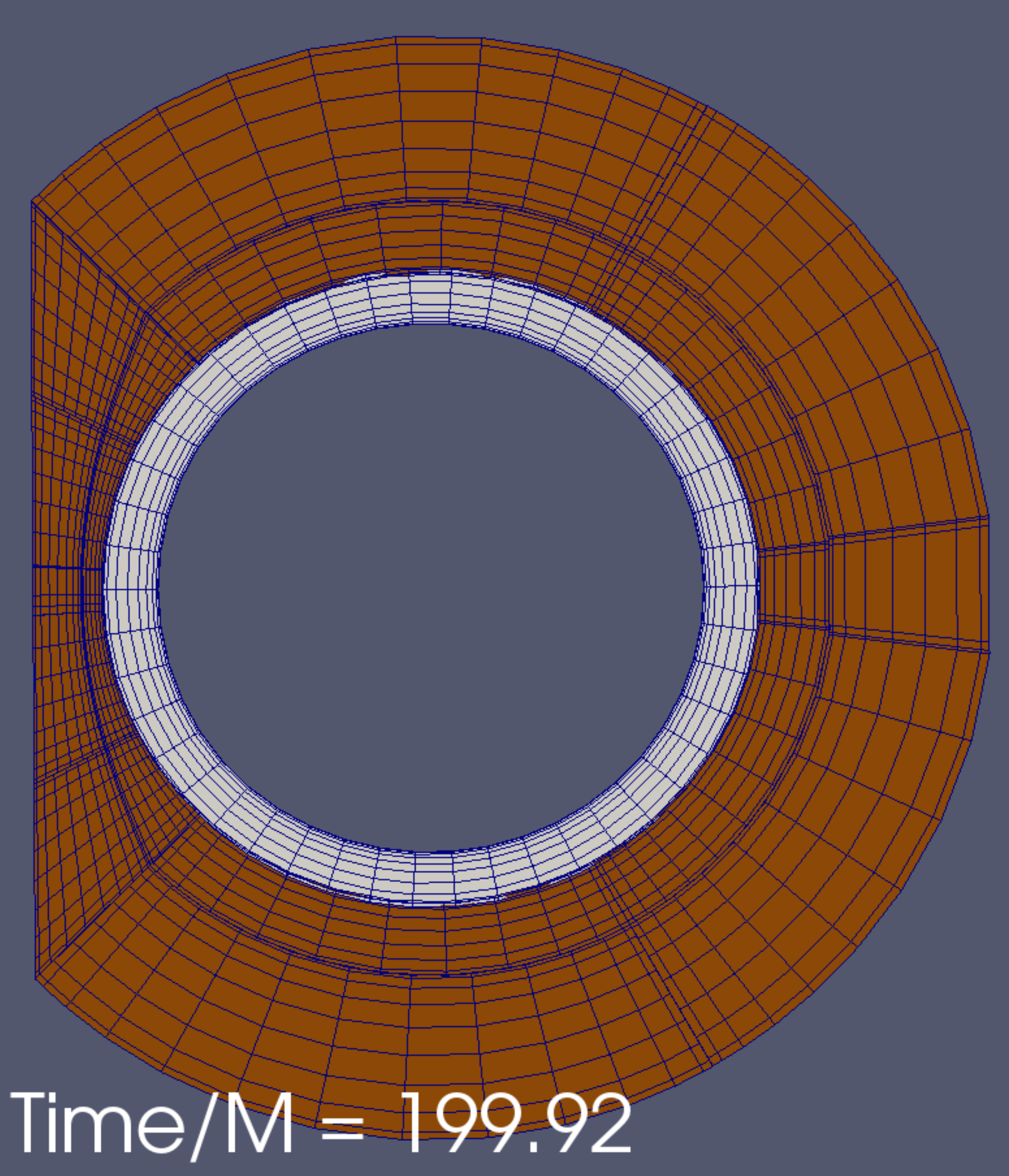}
}
\caption{\label{fig:ShellDropVtk}
Snapshots of the computational grid during the various stages of the shell dropping algorithm.
The { left panel} shows a $z=0$ cut of the grid structure around one of the black holes at $t=0$,
not including any of the outer cylinders or the wave-zone spherical shells.  At this initial
stage the black hole radius is small compared to the distance between the excision boundary
and the grid-plane on the left of the plot.  In this particular simulation  
the excision boundary is initially surrounded by nine spherical
shells.    The { middle panel} shows a similar cut,  at the point where six out of the nine shells have been
removed, as the excision boundary increases in size.  Note that the remaining three shells have radii
comparable to the outermost three shells of the initial configuration. 
The { right panel}
shows the grid when there is a single spherical shell left.  At this point the shell control
algorithm creates a thin spherical shell wide enough to fully contain the apparent horizon (as long
as this is feasible) but also thin enough to keep this last outer shell at a finite distance from the grid plane.
}
\end{figure}

As the black holes approach each other, the ratio of the size of the excision boundaries and 
the distance between their centers will grow.
In the co-moving frame, where this distance is fixed by definition, the signature of the plunge is that the apparent horizons 
(and the excision boundaries within them)
go through a rapid growth.    Our implementation of the shape map $\map{Shape}$ is such that the various
spherical shell boundaries around a given excision boundary stay at a constant distance from one another.   As the excision boundary
grows, this causes the outermost spherical shell to approach the plane cutting between the two sets of spherical shells,
as seen in Fig.~\ref{fig:ShellDrop}.
In order to prevent a grid singularity, when this condition is detected, we re-draw the computational domain,
typically reducing the number of spherical shells around the excision boundary by one. When we do this, we also
optimize the radii of the remaining shells so that they form a geometric sequence in the
co-rotating frame.  This is done several times as the black holes approach each other.  Near merger typically one would have a 
single spherical shell left around the excision boundary of the larger hole (see the right panel of Fig.~\ref{fig:ShellDropVtk}),
as this is quicker to approach the cutting plane.  
Especially for binaries with highly spinning black holes,
we find that it is important to keep at least one thin spherical shell around the excision boundary.  Not doing so (and thus having a set of distorted cylinders extend to the excision boundary) can introduce enough
numerical noise to cause our outflow boundary condition imposed at the excision boundary to become 
ill-posed.\footnote{At the 
excision boundary, by construction, we provide no boundary condition to any of the
evolution variables.  Numerically this means that we are evolving the innermost boundary points using 
a {\em sideways stencil}, i.e., using values from previously evolved points of the same subdomain, but no additional
boundary data.  This treament of the excision boundary is consistent with the requirements of the continuum system
as long as this inner boundary is spacelike.  If numerical noise becomes large enough, 
the inner boundary may become timelike on part of the excision boundary
and our numerical treatment is no longer adequate.}

\subsection{$r$-type mesh refinement}
\label{sec:rTypeAmr}

During the final stage of the plunge of high mass ratio BBH mergers the shell dropping
algorithm described in Sec.~\ref{sec:ShellDrop} does not provide sufficient control
of the grid.  This is due to the fact that our excision boundary can be very near the
individual apparent horizon surfaces and, as these approach each other, the excision
boundaries must also be able to get very near each other.  Given that in a typical simulation there
will be at least four subdomains placed between the excision boundaries and given
the number of maps connecting the grid frame with the inertial frame, each of which 
must stay non-singular (and invertible) during the entire simulation, one must
repeatedly redraw the grid in the immediate neighborhood of the excision boundaries
as the distance between these becomes a small fraction of their radii.

\begin{figure}
\centerline{
\includegraphics[scale=0.5]{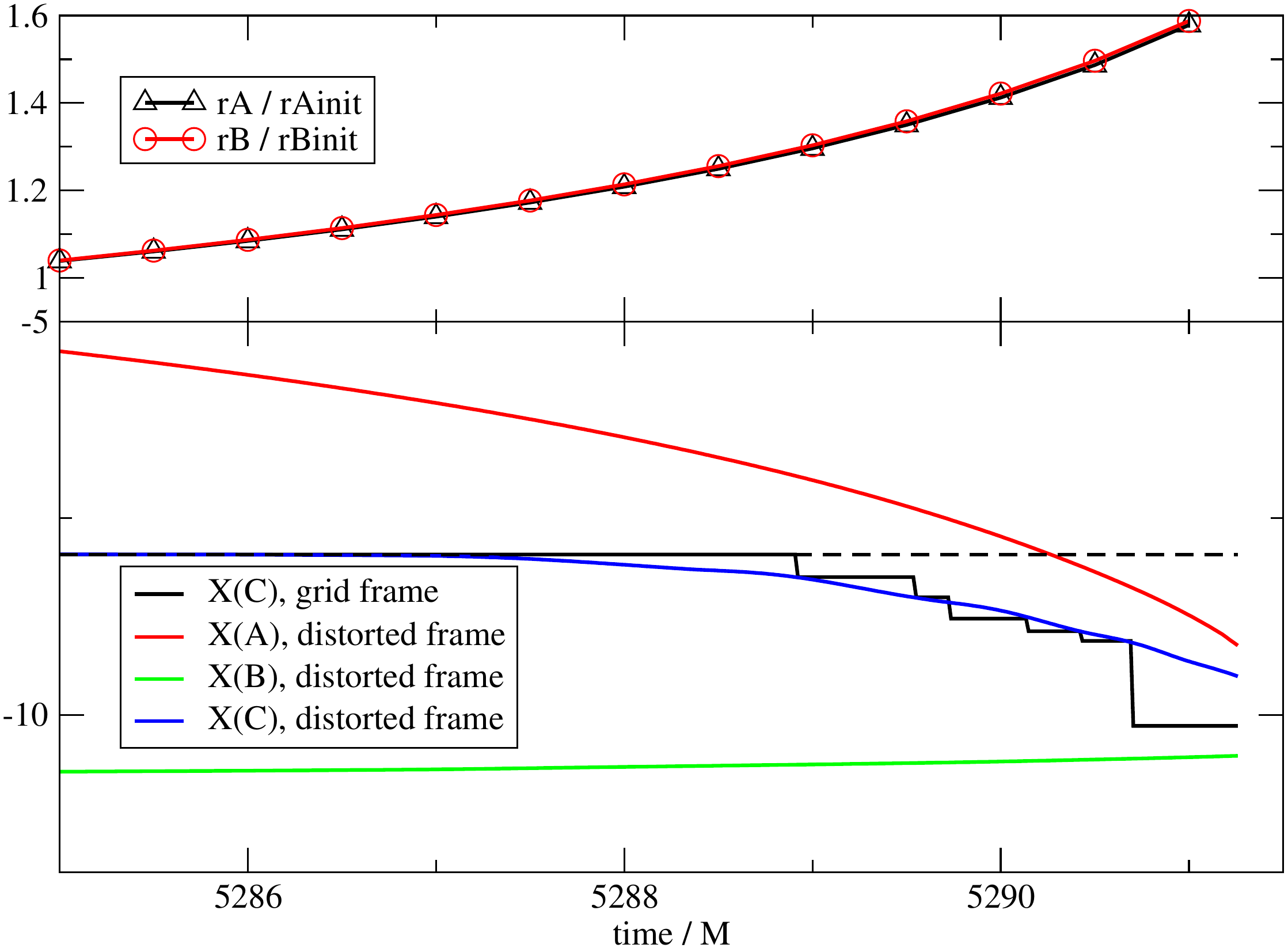}
}
\caption{\label{fig:CutXFuncOfTime}
  (Color online) { Upper panel:} The plot shows the average coordinate radius of the apparent horizon,
in the {\em distorted} frame, divided by the initial value of this average coordinate radius,
for a $q=9.98875$ non-spinning BBH simulation.  As the plot shows, the two black holes
grow at the same rate in this frame, although they have very different masses (and sizes).
This is consistent with the growth being a coordinate effect of $\map{Scale}$,
which shrinks the inner region of the grid as the black holes approach in order
to keep the excision boundaries concentric with them.  In this shrinking frame, the 
black holes have larger and larger coordinate radii, as seen in the plot.
{ Lower panel: }This plot illustrates how the location of the {\tt CutX} plane is controlled both in the
{\em grid} frame and in the {\em distorted} frame
for a $q=9.98875$ non-spinning BBH simulation. 
Let $S_A$ be the outermost spherical boundary around the larger black hole and
$S_B$ be the outermost spherical boundary around the smaller black hole.
The coordinate centers of both of these are near the $x$ axis in both the {\em grid} frame
and the {\em distorted} frame.  In the simulation, $S_A$ is to the right of the {\tt CutX} plane
(i.e., the $x$ coordinate value is larger than that of the {\tt CutX} plane for all of its points).
Similarly, $S_B$ lies left of the {\tt CutX} plane.  The green curve shows the $x$ coordinate
of the point of $S_B$ closest to the {\tt CutX} plane; similarly, the red curve shows the
$x$ coordinate of the point of $S_A$ closest to the {\tt CutX} plane.   As the plunge proceeds,
the black holes approach each other.  In the {\em distorted} frame this translates into
the green and the red curves approaching each other.  The task of the {\tt CutX} control system
is to smoothly move the {\tt CutX} plane out of the way of the black hole which approaches
it faster.  As it can be seen, without such a control system the code would have encountered
a grid singularity shortly after $t=5290$ (at the point where the red curve crosses
the dashed black line).   The {\tt CutX} grid-plane is moved out of the way
and the grid singularity is avoided.  As an additional requirement, the shape map 
$\map{Shape}$ must also not become singular during the simulation.  This is averted
by another algorithm that redefines the grid-frame location of the {\tt CutX}
plane, at discrete times.  This is displayed by the continuous black line. 
 Please refer to Fig.~\ref{fig:CutXSnapshots}
for further illustration of the effects of these discrete changes in the {\tt CutX} plane location.
}
\end{figure}

\begin{figure}
\centerline{
\includegraphics[scale=0.118]{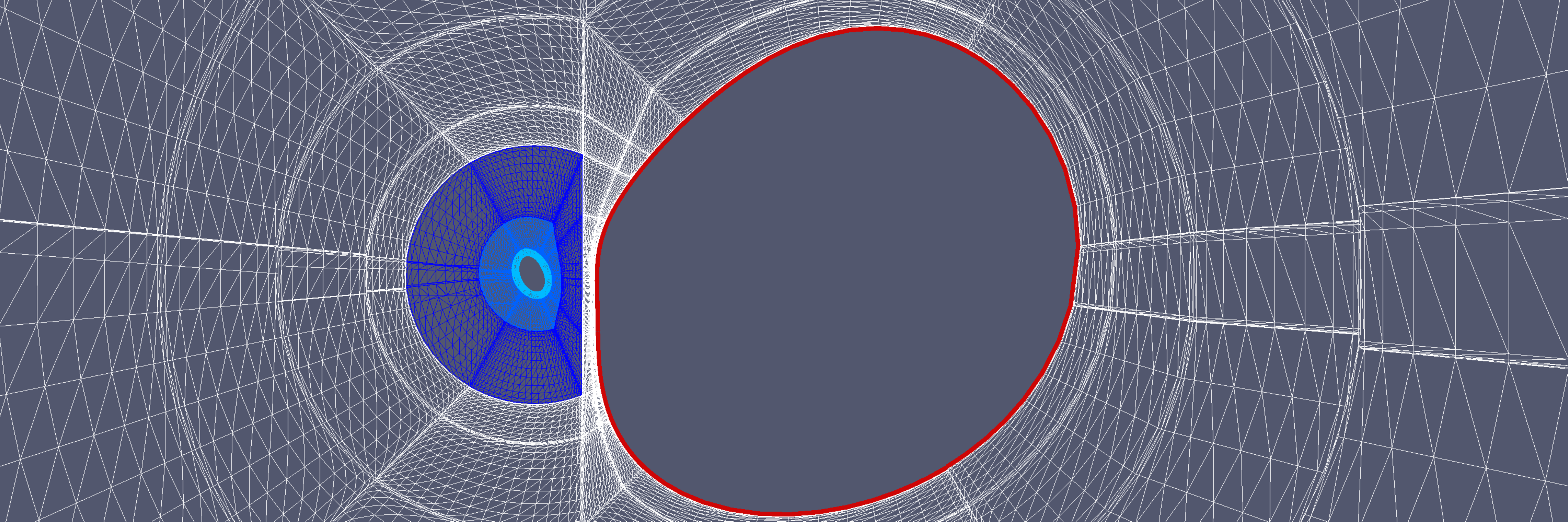}
\vspace{0.03in}
}
\centerline{
\includegraphics[scale=0.115]{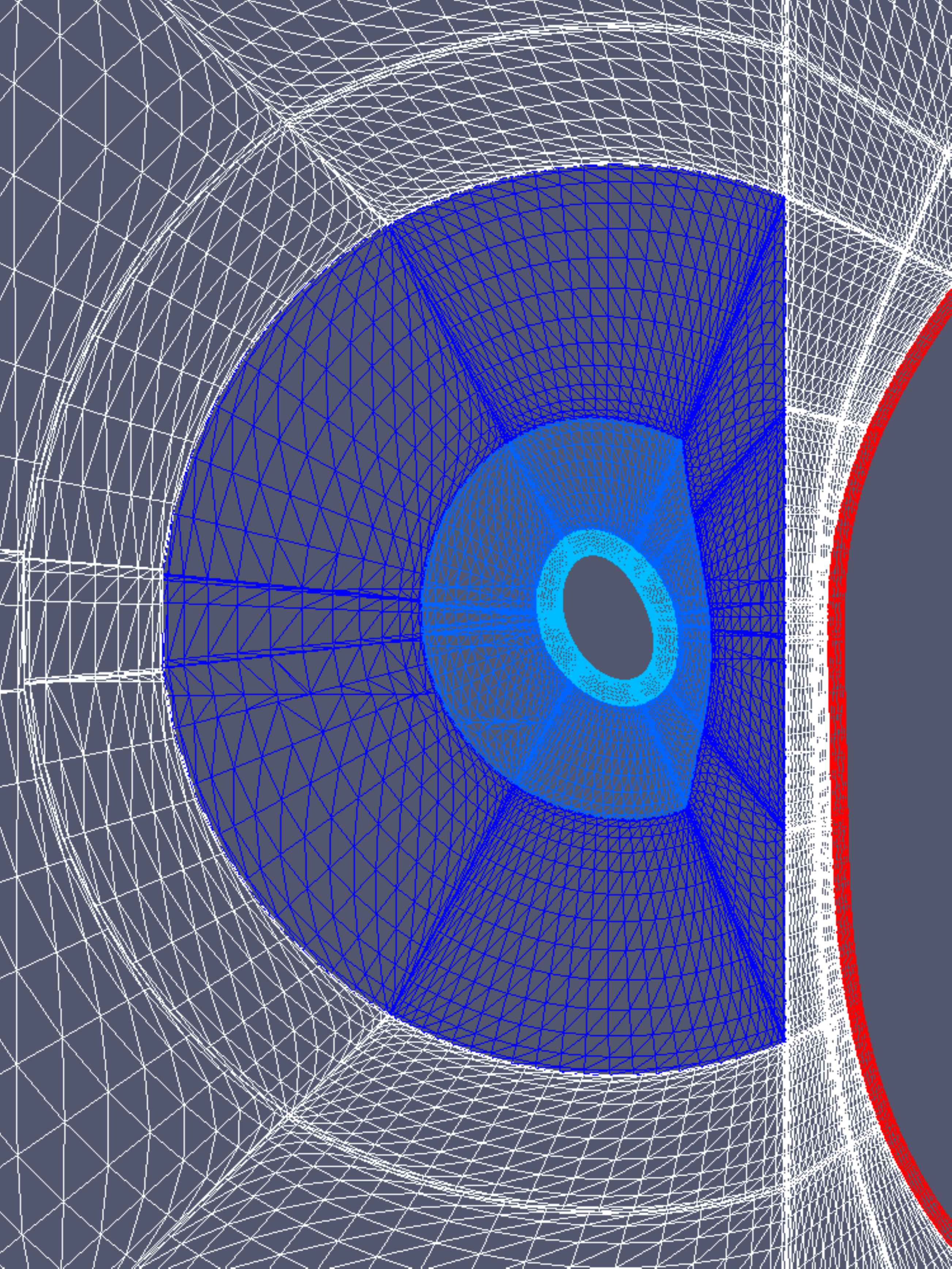}
\includegraphics[scale=0.115]{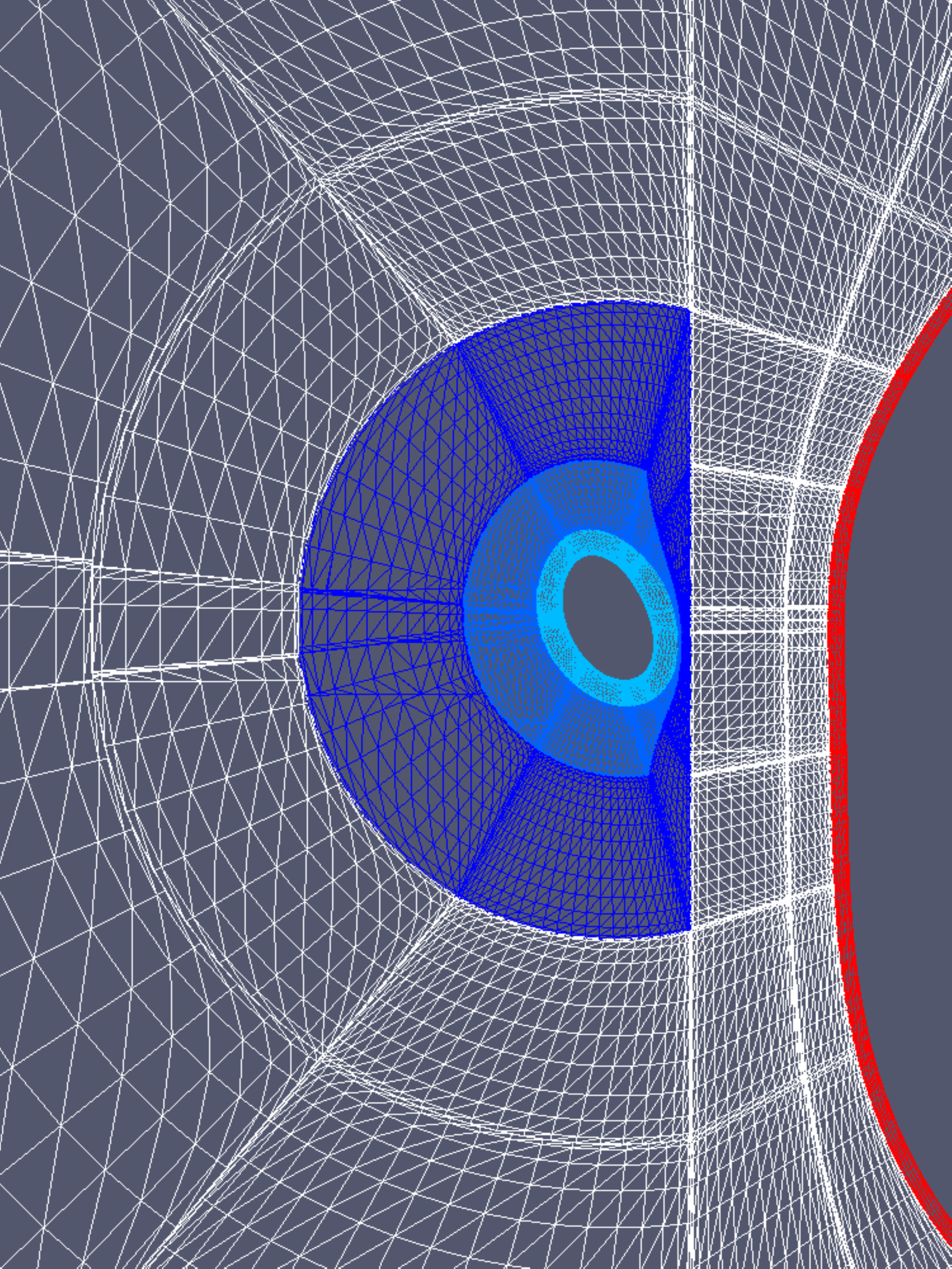}
\includegraphics[scale=0.115]{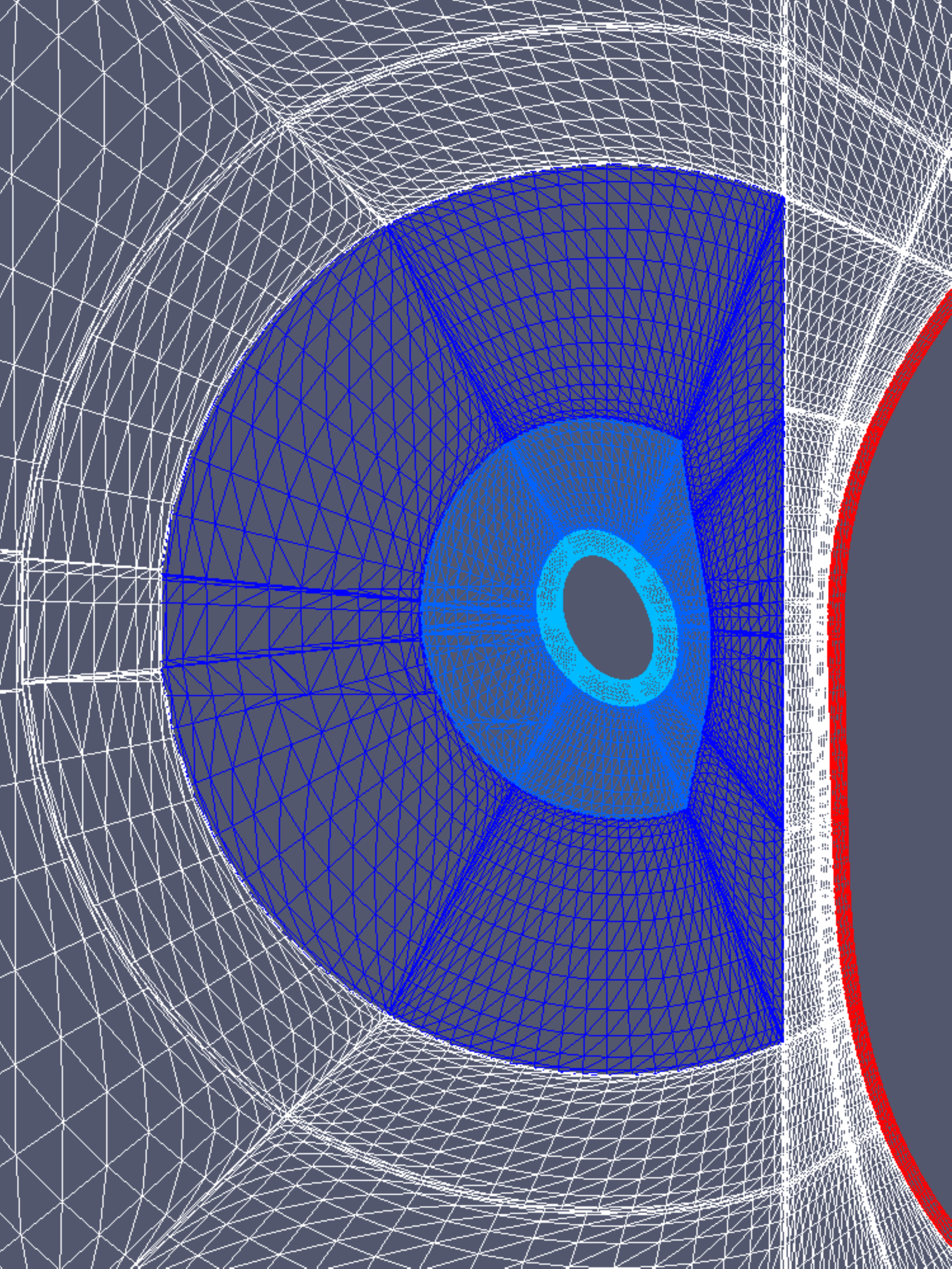}
\includegraphics[scale=0.115]{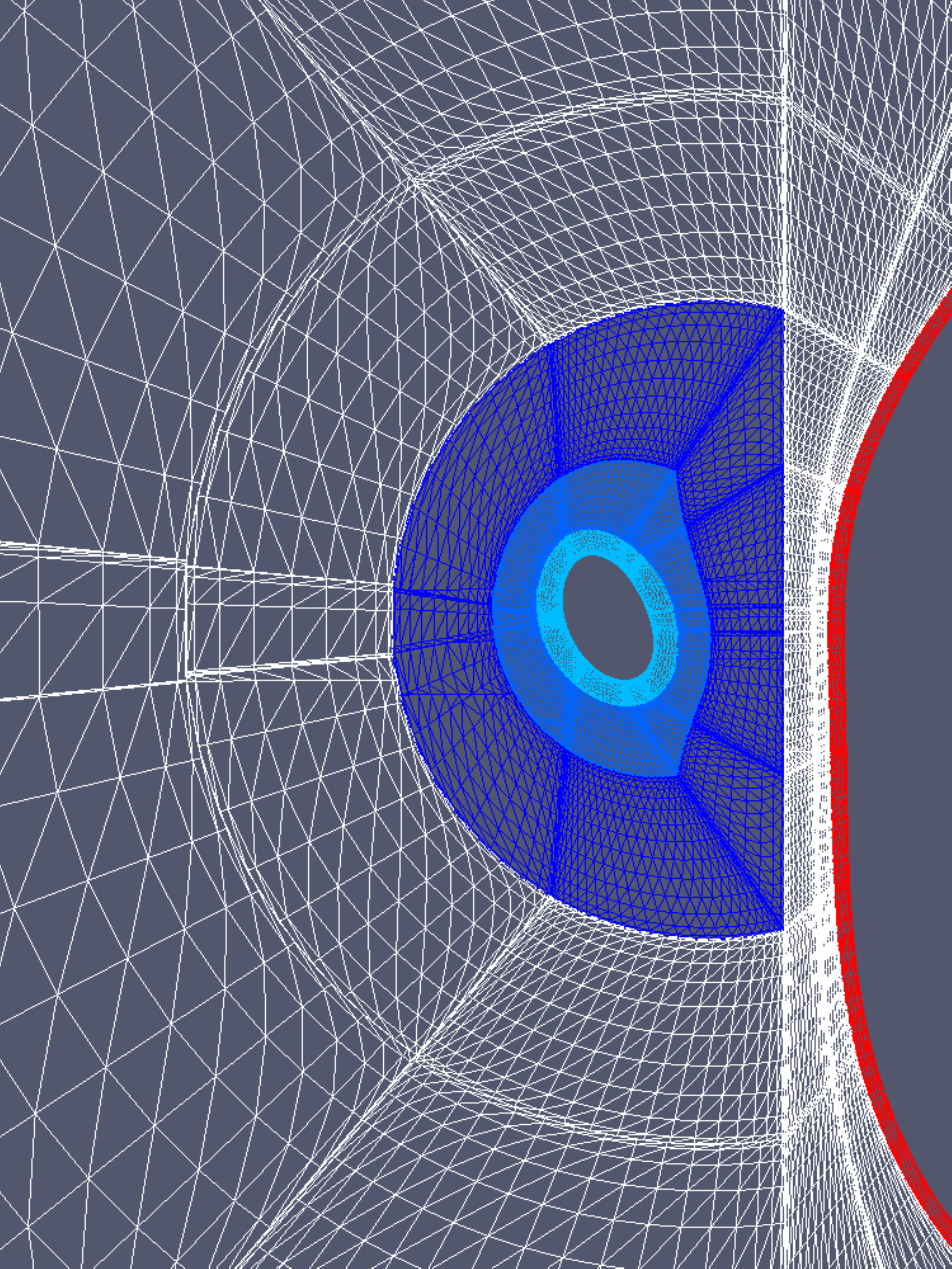}
}
\caption{\label{fig:CutXSnapshots}
  (Color online) Snapshots of the BBH grid, zoomed in around the smaller
black hole, for a 
$q=9.98875$ non-spinning BBH simulation, at $t\approx5290.69 M$.
The { top} plot shows the two excision boundaries and the surrounding grid-structure,
in the {\em distorted} frame.
An important feature of the grid is the {\tt CutX} plane, shown here as a vertical 
grid-line situated between the two excision boundaries.
The { lower first}
plot from left to right shows the grid in the {\em shape} map frame.   This map is
responsible for keeping the shape of the excision boundary in sync with that of the individual
black holes, while it leaves the {\tt CutX} plane unaffected. 
The { lower second} plot shows the grid
in the same frame, after {\tt CutX} plane position has been moved to the left, at a given time-instant.
One result of this grid change is that the excision boundary of the larger object has more room 
to grow.  Another effect of this grid change is that the subdomains in the immediate neighborhood
of the smaller excision boundary were reduced in size (see the blue region shrink from the first plot
to the second).  This gives additional accuracy to the region evolving the smaller black hole.
Note that the compressed grid between the excision boundary and the {\tt CutX} plane
does not factor into the CFL limit, as this is not the evolution frame.  The next map in the
sequence Eq.~(\ref{eq:MapSequence}), the $\map{CutX}$ map removes this compression, as 
seen on the fourth figure.  However, the extra room between the {\tt CutX} plane and the
(red) excision shell of the larger black hole is essential as this object grows at a fast
rate (see Fig.~\ref{fig:CutXFuncOfTime}).
The { lower third} plot is a zoomed-in version of the top plot.  
It shows the grid structure before the grid-change, in the {\em distorted} frame.
This frame differs from the {\em shape} frame by dynamically controlling the position of the {\tt CutX}
plane in order to keep it at a finite distance from the larger object as it rapidly expands.   The { lower fourth}
plot shows the {\em distorted} frame grid after the grid-frame position of the {\tt CutX} plane has been moved.
It is important to realize, in these plots, that the discrete relocation of the {\tt CutX} plane in the
 grid-frame does not affect (by construction) the position of the {\tt CutX} plane in the {\em distorted} frame.
Similarly, the smooth, time-dependent motion of the {\tt CutX} plane in the distorted frame does not imply
a change of the grid-frame location of this grid-plane.   The two are independent means of controlling
the shape of the grid in two different frames.  As one can see, the shrinking of the subdomains in the immediate
neighborhood of the smaller black hole can be also seen in the distorted frame (in the fourth plot).
}
\end{figure}

The one additional grid-boundary that needs to be dynamically controlled at plunge
is the {\tt CutX} plane.  As seen in Fig.~\ref{fig:CutXFuncOfTime},
the larger black hole tends to grow very rapidly in the distorted frame
in the final few $M$ of evolution time.   With its origin at a fixed location,
this implies that the $x$ coordinate position of its point closest to the small
black hole will rapidly approach the smaller object, with the {\tt CutX} plane in 
the way.  The dashed black line shows that if this plane were held at a fixed location,
the run would have crashed at $\sim 1M$ before merger.  The {\tt CutX} control
system described in Ref.~\refcite{Hemberger:2012jz} is designed to handle this, keeping the
map from the grid to the inertial frame non-singular throughout the merger.
However, as stated earlier,
it is essential that all individual maps in Eq.~(\ref{eq:MapSequence}) are
also non-singular and invertible (e.g., for the sake of the horizon finder
interpolator which needs to map the position of the horizon mesh points into
the grid frame).  This means that the excision boundary, under the action of the
shape map $\map{Shape}$,  must not cross the {\tt CutX} plane.  
For this reason, at a set of discrete times, we
re-position the {\tt CutX} grid-plane (as shown by the black curve in
Fig.~\ref{fig:CutXFuncOfTime}), such that it stays between the
excision boundaries (and the spherical shells surrounding these)
at all times, and in all frames. Fig.~\ref{fig:CutXSnapshots} shows
the grid-structure surrounding the two excision boundaries during
such a change.  Given that the subdomains are not split or joined but that
their shape is redefined, we regard this algorithm as $r-$type mesh refinement.

\subsection{Overall AMR performance}

\begin{figure}
\centerline{\includegraphics[scale=0.4]{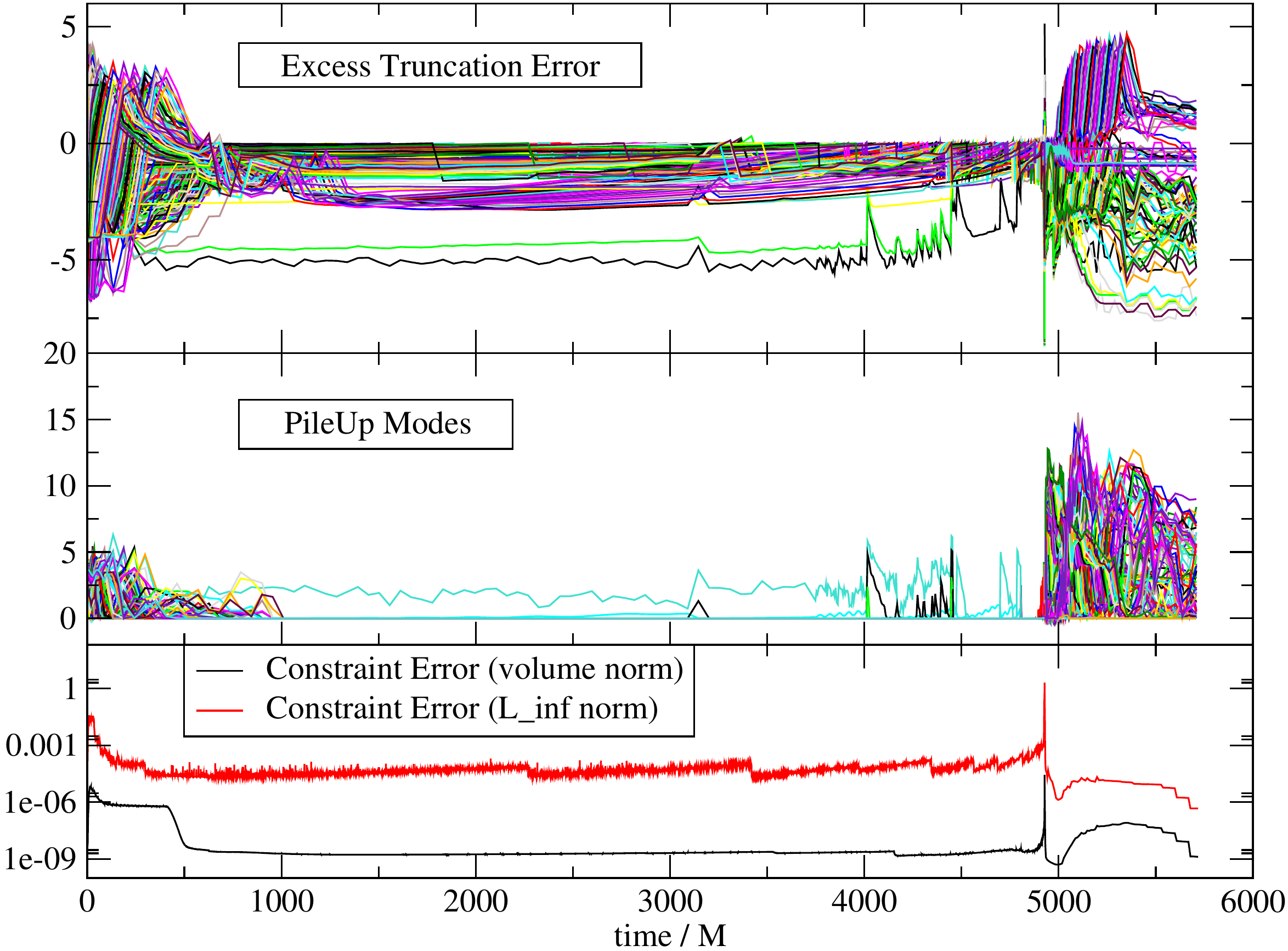}}
\caption{\label{fig:FullRunAmr}
Overall AMR driver performance for the  non-spinning $q=7.1875$ BBH evolution.
The { top panel} plots the `excess truncation error' defined as the difference between
the  truncation error Eq.~(\ref{eq:TruncError}) associated with the individual power-monitors $P_k$
and the target truncation error
Eq.~(\ref{eq:TruncErrorFunction}), ${\cal T}_{\rm excess} = {\cal T}[P_k] -{\cal T}^{\max}[w_A,w_B]$
for all spectral expansions in all subdomains.  The first $\sim 500 M$ of the simulation is dominated
by the `junk radiation' phase, where the black holes undergo quasi-normal ringing in response to
non-physical gravitational radiation content present in the initial data.  As the ringing subsides
and the high frequency waves leave the grid, the excess truncation error becomes negative, and
relatively close to zero, showing that the AMR driver is able to control truncation error.  Note that
various constraints on the subdomain extents 
(explained in Sec.~(\ref{seq:AmrRulesForShells})) imply that certain
subdomains have more accuracy than required.  The last $\sim 1100M$ of the run corresponds
to the ringdown part of the simulation.  As indicated in the plot, some of the modes (the radial modes)
are not well resolved, in a way similar to the initial junk radiation phase.  This is, once again, due to the
high frequency nature of the quasi-normal ringing of the final black hole.
This problem would benefit from further improvements of the AMR algorithm.
The { bottom} panel shows the $L_{\infty}$ and the $L_2$ norm of the (unnormalized)
constraint error.   This error is largest during the junk radiation phase except for a short peak
seen during the merger of the two black holes. This peak is no longer seen once the region inside
the common horizon is excised, suggesting that the larger constraint violating modes are
confined to the interior of the
merged black hole.
}
\end{figure}

As indicated in the various sections describing our AMR algorithm, a number of aspects
would benefit from improvement.  Overall, however, AMR has become
an essential part of our production simulations.  It avoids the need to repeat the same simulation,
time after time, iterating on the grid extents of the domain.   Neither do we have to contend 
with exponential error growth, a direct consequence of exponential convergence of our numerics
and the accelerated decrease of the length-scales, as dictated by the nature of the BBH 
inspiral and merger problem.   in Fig.~\ref{fig:FullRunAmr} we plot
the measured truncation error level for each irreducible topology
of each subdomain, the associated pile-up mode function and the overall constraint error,
as measured by the global $L_2$ and $L_\infty$ norms. This figure shows that during the
inspiral the AMR driver does quite well in keeping these quantities within control.  
The constraint error spike
shortly before $t=5000M$ corresponds to merger.   While one would prefer no such spike,
we note that the spike vanishes as soon as ringdown
part of our simulation begins, when the excision boundary is now based on the common horizon.  This
implies that the large errors are well within the newly formed black hole and do not
affect the physics of the region outside the new excision boundary.

Aside from this spike, there are two other problem areas one can point to on this plot.  The {\em junk radiation}
has been a long-standing problem for us.  Resolving it would require an unreasonable amount of
computational resources for a given simulation.   We are looking for alternative ways of tackling this problem.
Another problem arises in the ringdown, where truncation reaches levels near $10^{-5}$
and pile-up modes are created.   The constraint error in this part of the run remains small, however.
One possible reason why our truncation error estimate indicates large error is that the grid used
to propagate the gravitational waves generated by the quasinormal modes may not be optimal
once the black hole settles down.  A proper $h$-type AMR driver would possibly join the rather large
number of spherical shells into a handful of subdomains once the spacetime reaches a 
smooth, stationary state. 

Our largest concern, at the moment, is improved junk radiation treatment, as noise in the late ringdown
has only a minimal effect on the primary output of our code -- the gravitational waveform emitted
by the BBH system.

\section{Optimization}

\begin{figure}
\centerline{
\includegraphics[scale=0.2]{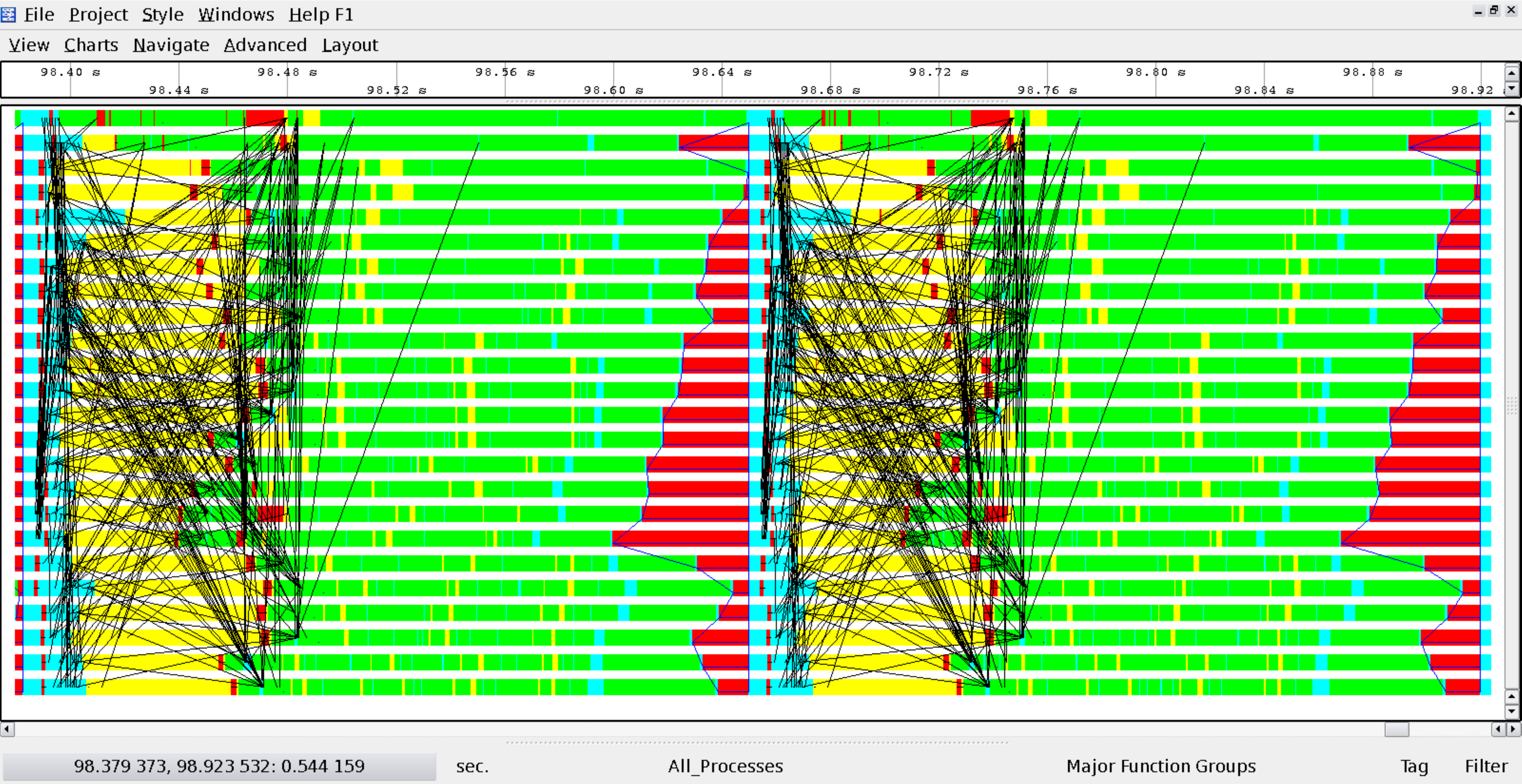}
}
\caption{\label{fig:PID} (Color online)
The plot illustrates the inner workings of our load balancing algorithm.
This output is produced by the {\em Intel Trace Analyzer}, measuring
performance of our code for two sub-time-steps of a typical BBH evolution.
The horizontal direction corresponds to wall-clock time, ranging from $98.40$sec
to $98.92$sec, measured from the beginning of the test-run.  Each of the horizontal
stripes represents an {\tt MPI} process.  Black lines represent {\tt MPI} messages sent from one
process to another.  The green portions correspond to computation of the various parts
of the evolution system's RHS (in this case Eq.~(\ref{eq:RicciH})).
Yellow portions indicate differentiation of the evolution variables, while cyan stands
for interpolation of boundary data at inter-subdomain boundaries.  Red stands for {\tt MPI}
communication.  At the end of each sub-time-step an {\tt MPI\_Barrier} is placed to force syncing
between the various processes.  The cost of this barrier on the individual processes is used as an
indicator of the local process load.  If the barrier cost is minimal, the processor is busy and 
is among the last to reach the barrier.  At subsequent RHS evaluations work load is
shifted away from this process.   In this simulation the processor with rank zero has
the largest load.  By implication nearly all differentiation and interpolation
task has been shifted away from this process to others with a smaller workload.   As an additional
feature, all {\tt MPI} messages sent during the evaluation of the RHS are non-blocking (or
sent in the background).   The RHS evaluation algorithm then periodically calls back to the
parallel adaptive interpolation/differentiator algorithm to test whether any of the non-blocking
messages have reached their destination.  If yes, the data is processed (e.g., differentiated)
and then sent on its way in another non-blocking message.  At the end of the RHS evaluation,
when a given piece of the data is required for further work, an {\tt MPI\_Wait} is issued for the
required message.  In most cases, however, the messages will have arrived by this time and
there is minimal communication cost (shown in  red) during RHS evaluation.
}

\end{figure}

As a last element of our description we highlight one particular aspect of the optimization
work that was done to {\tt SpEC}.    In our current parallelization scheme tasks associated with
a particular subdomain are handled by a single process, with no multi-threading.   Given the small
number of such subdomains, the maximum number of processes one can meaningfully
use in a BBH simulation is $48$ (assuming $16$ or $12$ core nodes).   This on its own is a limitation
we'll possibly need to deal with by adding multi-threading to our code.  But, even before that,
the varying number of points per subdomain means that if each process owns a single subdomain,
the work load per process would be far from even.  Reducing the number of processes would
help a little but, even in that case, load balancing would be very coarse grained.   In order to improve this
situation, we have implemented an adaptive parallel interpolation and differentiator (PID) algorithm that
dynamically re-assigns as much as $30\%$ of the workload associated with a given subdomain 
from  processes  with too much work to those with too little.   {\tt MPI\_Barrier} calls, issued
after each RHS evaluation, are used to monitor which process is working
hardest.  If the cost of these {\tt MPI\_Barrier} calls is consistently much smaller on a particular
process than on some other process, the process with the small measured   {\tt MPI\_Barrier}
cost must be busy (see Fig.~\ref{fig:PID}).   In this case, the process would no longer be
responsible for interpolating data, as required in the inter-subdomain boundary algorithm.
Neither would this process be responsible for differentiating the evolution variables. Rather,
through a sequence of non-blocking {\tt MPI} calls, this data would be sent to a less busy process.
After all such initial messages have been sent, each process starts to compute
the various pieces of the RHS (in this case: the pieces of Eq.~(\ref{eq:RicciH}) that do not involve
derivatives).   At various points during this RHS evaluation, the evolution algorithm dials back
into the adaptive PID algorithm to see if any of the non-blocking messages have arrived at their
destination.  If so, data is processed (i.e., interpolated and/or differentiated).  Then the processed
data is sent on its way -- the derivative data needs to get back to the process responsible for updating
the corresponding subdomain, while the interpolated data is sent to those processes that need this
data to complete the RHS evaluation on the subdomain boundary points.  At some point in the RHS
evaluation the derivative data or boundary data becomes a {\em must}, i.e., no more work that can
be done without it.   At this point blocking {\tt MPI\_Wait} calls are issued, enforcing the arrival
of all data needed to complete the RHS.   At seen on Fig.~\ref{fig:PID}, most communication is
in fact complete during the first half of the RHS evaluation and there is virtually no waiting for data.

This, as well as other optimization techniques, have lead to  significant performance improvement of
the {\tt SpEC} code.  As stated earlier, there is more work to be done, in particular in the area
of multi-threading.

\section{Conclusion}

This paper gives a tour of the main building blocks of a state-of-the art binary black hole
evolution code, known as the {\tt SpEC} code.   We start by writing out the evolution system
used in our  simulations, a first order symmetric hyperbolic form of the
 generalized harmonic formulation.  We then provide a brief description of the numerical
approximation used in the code.  Based on these concepts,  we then
detail the diagnostic quantities used to monitor (and control) the accuracy of our simulations.
The definitions we provide for the  truncation error, the pile-up modes and the convergence factor
are then used as
we give an elaborate description of our spectral adaptive mesh refinement algorithm, tuned
to the problem of binary black hole evolutions.   As laid out in our description, our favored
method of controlling truncation error is $p$-type mesh refinement, with a target truncation
error function that is tuned to use more resolution in the strong field region.  In addition, we
use $h$-type mesh refinement mostly as a means of preventing grid compression as the size
of the apparent horizons (and the associated excision boundaries) increases relative to 
the distance between the coordinate centers of the binary.  Late in the plunge, for
binaries with sufficiently large mass ratios, we find $r$-type mesh refinement to be essential;
as the black holes approach, the grid layout defined at the beginning of the simulation
is no longer adequate.  We update the grid boundaries as dictated by the dynamics of the binary 
black hole plunge and merger.

Fig.~\ref{fig:FullRunAmr} illustrates how well our current algorithm performs during most of the inspiral.
However, our algorithm does not do well at the early stage of the run where the unphysical gravitational
wave content of the initial data induces high frequency ringing of the individual black holes, which in turn
generates short wavelength gravitational waves which propagate through a grid
that is not designed to resolve them.
As a consequence of not resolving this ``junk radiation'', there is a stochastic part of the
truncation error whose noise produces a floor in our convergence measurements.
Our attempts to control this stochastic noise source is focused on
improving the initial data algorithm, e.g., by use of the `joint-elimination method' described
in Ref.~(\refcite{Zhang:2013gda}).   

Another phase of our run where the AMR algorithm
may need improvement is the ringdown stage.  The reason behind this is mostly historic -- this has been the easiest
part of the simulation, as evolving a single black hole is simpler than evolving a binary.
We plan to improve this part of our simulation
but presently it is not a the major limitation in the
quality of our gravitational waveforms.

Lastly, we have provided a brief description of part of the optimization that was done on the code.  This is
very ``raw'' computer science, but it is an essential part of why are we able to perform inspirals for dozens of 
(or, in once case, nearly two hundred) orbits.

Future work will focus on improving the accuracy of the phase of the gravitational wave during the plunge.
The dominating error source in this very dynamic 
part of the simulation appears to be a sub-optimal structure of the
inner grid at the point where nearly all spherical shells have been removed by our shell control algorithm.  This results in 
accumulation of orbital phase error at a faster rate than during any other part of the simulation. We expect
to improve this by refining our $h$-type AMR algorithm. We also plan on exploring the 
use of the constraint error driven target
truncation error, as this will tighten accuracy requirements when the constraints start to grow.

As yet another future project, we plan to improve our outer boundary algorithm, as 
we find that reflections from the outer boundary can cause unphysical
effects on the binary evolution on very long time-scales (beyond $100$ orbits).   

The end goal is to enable the construction of 
semi-analytic (or empirical) gravitational wave models and template banks by providing 
numerical gravitational waveforms
of sufficient accuracy, length and quantity such that  modeling error does not detract from
the detection of gravitational waves, or the resulting physics one can discern from the 
observed data.

\section*{Acknowledgments}

We thank Jeffrey Winicour for his useful comments on the manuscript.   We thank Dan Hemberger for
providing access to data produced by the head-on BBH simulation used in some of the plots.  We thank
Jonathan Blackman for providing access to his unequal mass inspiral simulations, as these provided
data for most of the plots.  We thank Saul Teukolsky, Mark Scheel, Larry Kidder, Harald Pfeiffer and Nicholas Taylor for
helpful discussions while designing the AMR code.  We thank Abdul Mroue, Tony Chu, Geoffrey Lovelace and Sergei Ossokine
 for useful feedback
on the performance of our AMR algorithm in the context of thousands of BBH simulations.
A significant amount of code development was done on the UCSD cluster
{\tt ccom-boom.uscd.edu}.  The Caltech cluster {\tt zwicky.cacr.caltech.edu} is an essential element
of research done with {\tt SpEC}.  This cluster is supported by the Sherman Fairchild Foundation and by NSF
award PHY-0960291.  This project was supported by the Fairchild Foundation,
and the NSF grants  PHY-1068881 and  AST-1333520.

\bibliographystyle{ieeetr}
\bibliography{References/References}

\end{document}